\documentclass[english,12pt,sort&compress]{iopart}
\usepackage[T1]{fontenc}
\usepackage{float}
\usepackage{iopams}
\usepackage{graphicx}
\usepackage{amssymb}
\expandafter\let\csname equation*\endcsname\relax
\expandafter\let\csname endequation*\endcsname\relax
\usepackage{amsmath}
\usepackage[numbers]{natbib}
\usepackage{verbatim}
\usepackage{color}


\newcommand{\A}[0]{\mathcal{A}}
\newcommand{\B}[0]{\mathcal{B}}
\newcommand{\HH}[0]{\mathcal{H}}
\newcommand{\M}[0]{\mathcal{M}}
\newcommand{\sig}[0]{\boldsymbol{\sigma}}
\usepackage{babel}

\makeatother

\begin{document}

\title[]{Shared information in classical mean-field models}
\author{O Cohen$^1$, V  Rittenberg$^2$ and T Sadhu$^3$}

\address{$^1$ Department of Physics of Complex Systems, Weizmann Institute of
Science, Rehovot 76100, Israel.}

\address{$^2$ Physikalisches Institut, Universitat Bonn, Nussallee 12, 53115
Bonn, Germany.}

\address{$^3$ Institut de Physique Th\'{e}orique, CEA/Saclay, Gif-sur-Yvette Cedex, France.}

\ead{\mailto{or.cohen@weizmann.ac.il}}

\begin{abstract}
Universal scaling of entanglement estimators of critical quantum systems has drawn a lot of attention in the past. Recent studies indicate that similar
universal properties can be found for bipartite information estimators of classical systems near phase transitions, opening
a new direction in the study of critical phenomena. We explore this subject by studying the
information estimators of classical spin chains with general mean-field interactions.
In our explicit analysis of two different bipartite
information estimators in the canonical ensemble we find that, away from
criticality both the estimators remain finite in the thermodynamic limit. On the other hand, along the critical line there is a
 logarithmic divergence with increasing system-size.
The coefficient of the logarithm is fully determined by the mean-field interaction and it is the same for the class of models we consider.
The scaling function, however, depends on the details of each model.
In addition, we study the information estimators in the
micro-canonical ensemble, where they are shown to exhibit a different universal behavior.
We verify our results using numerical calculations of two specific cases of the general Hamiltonian.
\end{abstract}


\vspace{2pc}

\submitto{\JPA }

\maketitle

\section{Introduction}
The study of entanglement properties in quantum many-body systems
has attracted considerable attention in recent years (see \cite{Eisert2010} for a review).
The measures of entanglement provide a promising tool for understanding universal
properties of quantum systems, in particular, at the vicinity of quantum phase transition.
Typically, as a simple estimation, the entanglement is measured between two fictitiously
partitioned subsystems $\A$ and $\B$ in the ground state $\vert \Psi_{0}\rangle $ of the whole
system. A quantitative measure of this bipartition entanglement is the von-Neumann entropy of the reduced density matrix
$\rho_\A=\tr_{\B}\rho$, defined as
\begin{equation}
\mathcal{E}_{\A:\B}=-\tr_{\A}\rho_{\A}\ln \rho_{\A},
\end{equation}
where $\rho=\vert \Psi_0 \rangle\langle \Psi_0 \vert$ is the density matrix of the entire system.

This entanglement estimator has been widely studied in several quantum systems
\cite{Holzhey1994,Vidal2003,Korepin2004,Gioev2006,Li2008,Calabrese2008,Alba2012}.
Although the quantity appears extensive, it is typically found to be
proportional to the area of the hyper-surface separating the two
subsystems, particularly when the system is in a gaped phase. This is the
celebrated area law \cite{Srednicki1993,Wolf2008,Eisert2010}. What is more
interesting is that,
when the system is critical (or gap-less) there is correction to this area law. Moreover, the correction has universal properties.
For a one dimensional quantum system which exhibits a conformal symmetry, it was shown that the entanglement estimator $\mathcal{E}_{\A:\B}$
diverges logarithmically with the increasing system length $L$ \cite{Holzhey1994,Calabrese2004}. Moreover, the divergence obeys the following scaling form:
\begin{equation}
\label{eq:area_div}
\mathcal{E}_{\A:\B}=\frac{c}{3}\ln \left[ L \psi\left( \frac{\ell}{L}
\right)\right]+\textrm{constant},
\end{equation}
where $\ell$ is the size of the subsystem $\A$, and $\psi$ is the scaling
function. The additive constant in \eref{eq:area_div} is non-universal in the sense that it depends on the
microscopic details of the model. Remarkably, the constant $c$ turns out to be
universal. For periodic systems, $c$ is equal to the central charge of the underlying conformal field
theory whereas for open systems it is equal to half of the central charge \cite{Holzhey1994,Vidal2003,Calabrese2004}.
For critical quantum systems, the central charge characterizes the long-distance physics of the system. For example, the central charge of the quantum Ising system is equal to $1/2$.

Our understanding is less complete for higher-dimensional systems.
The area law has been generally proven in the gaped phases of a
systems with short-range interactions \cite{Wolf2008}. At criticality there are mixed examples: in some systems, such as free bosons, the area law
is found to be satisfied \cite{Plenio2005,Cramer2006,Calabrese2012}, whereas in
other systems, such as free fermions, there are logarithmic corrections to the area law \cite{Song2011,Barthel2006,Wolf2006,Gioev2006,Swingle2010}.

There are also other bipartite estimators of entanglement such as the R\'enyi entropy
\cite{Calabrese2010}, the mutual information \cite{Um2012,Alcaraz2013}, the
quantum discord \cite{Ollivier2001}, the logarithmic negativity \cite{zyczkowski1998volume,eisert1999comparison}, which
exhibit similar universal behavior \cite{Korepin2004,Sarandy2009}.

Naively, all these estimators measure the information shared between the degrees of
freedom in the two subsystems.
A natural question to ask is whether in classical systems information estimators, similar
to the entanglement entropy, exhibit an area law. Moreover, it is interesting to study whether such estimators
exhibit logarithmic corrections to the area law at criticality, in a manner that characterizes the universality class of the classical system.
The first question has been addressed by Wolf \textit{et al.} \cite{Wolf2008}, who
studied an estimator known as the mutual information.
It quantifies the amount of information acquired about the configuration of
one subsystem by measuring the state of the other. They have shown that the mutual information
of any classical system with a finite correlation length obeys an area-law.
The second question, about criticality, has recently been addressed by Alcaraz and Rittenberg \cite{Alcaraz2010}, who
studied the scaling of the mutual information as well as other information
estimators for several classical stochastic models. They have shown that at
criticality the estimators exhibit logarithmic corrections to the area-law,
with a scaling similar to \eref{eq:area_div}, observed in quantum systems.
The coefficient $c$ was found to depend on the model and the estimator studied.
However, for a specific model and estimator, $c$ was found to be independent of the parameters of the model
and thus remains constant along the critical line.  This universal behavior suggests that as in the quantum case, the
coefficients $c$ could be useful in characterizing the universal properties of classical many-body systems.

Unfortunately, there is almost no other example where the scaling properties of the
shared-information estimators have been studied in classical many-body systems. In some cases,
the mutual information has been proposed as a means to detect phase transition in classical
spin models, where it was shown numerically to exhibit non-analytic behaviour
\cite{Iaconis2013,Lau2013}. However, an analytical computation of the mutual information
estimator is often very difficult.

In this paper, we address the issue by studying shared-information estimators
in classical spin-chains with mean-field interactions of a general form.
Due to the long-range nature of their interactions such models exhibit non-trivial phase diagrams.
At the same time they are simple enough for detailed analytic calculations.
Our goal is twofold: first, to study different estimators and compare their behavior
across the phase diagram, particularly along the critical line. Second, to study how the scaling behavior changes from
one thermodynamic ensemble to the other, in particular from the canonical to the
micro-canonical ensemble. We study two shared-information estimators: the mutual information $(\mathcal{I}_{\A:\B})$ and separation entropy
$(S_{\A:\B})$.

By carrying out an explicit analytical calculation, we find that within the canonical
ensemble and away from criticality, both the estimators remain finite as the system length tends to infinity.
 This is not obvious a priori for systems with long-range interactions. At criticality, we find a different scenario. The
mutual information exhibits a logarithmic divergence similar to
\eref{eq:area_div} with $c=1/4$, with $\ell$ and $L-\ell$ denoting the number of spins in the two partitions. On the other hand, the separation
entropy has a $\sqrt{L}$-divergence in addition to the $\ln L$ term.
In both estimators, the coefficient of the logarithmic term does not depend on the
microscopic details of the model, as in the quantum case. On the hand, the scaling function $\psi$ does depend on the details of model.
We also demonstrate that
for both estimators the coefficient of $\ln L$ remains the same even in the presence of additional
short-range interactions.
 This suggests that the value of this coefficient is characteristic of the mean-field universality class.
Unlike in the entanglement entropy used for quantum systems, the coefficient does not depend on the number of states each spin takes.

It is important to note that when considering only mean-field interactions the notion of geometry is lost, and thus the
 area-law is not well defined. The fact that the estimators remain finite in the thermodynamic limit, can be considered as equivalent to the area-law of one dimensional systems.

 The spin-spin correlation in mean-field models,
$c(r)=\langle\sigma_{i}\sigma_{i+r}\rangle-\langle\sigma_{i}\rangle\langle\sigma_{i+r}\rangle$
does not depend on $r$ because all spins interact with all the other spins. The critical point in these models is characterized by a change in the scaling of $c(r)$ from $c(r) \sim 1/L$ away from criticality to $c(r) \sim 1/\sqrt{L}$ at criticality. This is reflected in the divergence of the information estimators at the critical point.

In the micro-canonical ensemble, where the energy is strictly conserved, the
estimators exhibit a very different behavior. In our analysis we find that
the fixed-energy constraint imposes additional correlations between the local
degrees of freedom of the subsystems, which result in an additional $(1/2)\ln L$ terms
in both the estimators. As a result, even away from criticality we find a logarithmic divergence similar to \eref{eq:area_div}. At criticality,
the mutual information scales as $(3/4)\ln L$ while the separation entropy
scales as $(1/4) \ln L$. Notably, the leading $\sqrt{L}$ scaling
seen in separation entropy in the canonical ensemble is absent in the
micro-canonical ensemble. In our detailed analysis we show that this term is
associated to the fluctuations of the total energy which are absent in the micro-canonical ensemble.

To test our analysis we compare with numerical results of two particular
realizations of the general Hamiltonian, which we studied analytically: one is the mean-field
variant of the Blume-Emery-Griffiths model \cite{Blume1971,Barr2001} and the second is the
Nagel-Kardar model \cite{Nagle1970,Bonner1971,Kardar1983,Mukamel2005}. The former is a
$3$-state spin model with pure mean-fields interactions.
The latter is an Ising model with additional mean-field interaction. Both models have been
 studied extensively in the past, serving as prototypical models of the long-range
interacting systems.  Results from our numerical analysis of these two models is found to be in good agreement with
our analytical results. Another instance of our generic Hamiltonian is the
Curie-Weiss model. This has been studied analytically by Wilms \textit{et al.}, who computed the mutual information within the canonical ensemble \cite{Wilms2012}. In our analysis we recover their results.

The layout of this paper is as follows. \Sref{sec:background} provides the
background for our theoretical analysis, introducing the information estimators
and a brief description of the models considered.
The main results of our study are summarized in \sref{sec:main}. A
detailed analysis of the information estimators are then presented for the
generic model in \sref{sec:generic_model}. Concluding remarks are given in \sref{sec:conc}.

\section{Background}
\label{sec:background}
\subsection{Estimators of shared information in classical spin chains}
\label{sec:estimators}
The idea to measure mutual information between two random variables was first
introduced by Shannon in the context of the theory of communication \cite{Shannon}.
In recent years this approach has been extended to systems with many degrees of freedom. While a measure
of the information among all variables in a system (multipartite information) is hard to compute, we can learn much from measuring the
mutual information between two macroscopic parts of the system (bipartite information). It is possible to define
more than one estimator of shared information in bipartite systems, as demonstrated in \cite{Alcaraz2010}.
They all measure in different ways the mutual dependence between two compartments of a system and quantify
the amount of uncertainty about one subsystem when knowing only the state of the other.
In this paper, we study two such bipartite information estimators, namely, the separation entropy estimator ($S_{\A:\B}$), the mutual-information estimator
($\mathcal{I}_{\A:\B}$).

We define the estimators for a classical spin chain of size $L$,
 where every site is occupied by a spin variable $\sigma_{i}$
that takes $p$ discrete values, $\sigma_i=1,\dots,p$. We consider a spatial bipartition of the
system into two parts, $\A$ and $\B$, of size $\ell$ and $(L-\ell)$,
respectively, such that sites $\{1,...,\ell\}$ belong to subsystem
$\A$ and the remainder to subsystem $\B$. A configuration of the system is denoted
by $({\sig}^{\A},{\sig}^{\B})$, where ${\sig}^{\A}\equiv\{\sigma_{1},\dots,\sigma_{\ell}\}$
and ${\sig}^{\B}\equiv\{\sigma_{\ell+1},\dots,\sigma_{L}\}$
are the spin configurations of the two subsystems. We denote the equilibrium probability of a configuration by $P\left({\sig}^{\A},{\sig}^{\B}\right)$.
In order to define the information estimators one has to consider also the marginal probability distribution of each subsystem,
obtained by summing over the configuration of the other subsystem, yielding
\begin{equation}
P_\M^\A({\sig}^{\A})=\sum_{{\sig}^{\B}}P({\sig}^{\A},{\sig}^{\B}),\qquad
\textrm{and} \qquad P_\M^\B({\sig}^{\B})=\sum_{{\sig}^{\A}}P({\sig}^{\A},{\sig}^{\B}).
\end{equation}
In addition, we consider the probability distributions of the two subsystems when they
are physically decoupled, denoted by $P(\sig^{\A})$ and $P(\sig^{\B})$.
The decoupling is obtained by turning off all the interactions between spins belonging to different subsystems.
Note that in the decoupled state, the distribution of the composite system is given by a product form,
$P(\sig^{\A},\sig^{\B})=P(\sig^{\A})P(\sig^{\B})$,
 yielding $P^{A}_\M(\sig^{A})=P(\sig^{A})$ and $P^{B}_\M(\sig^{B})=P(\sig^{B})$.

The two information estimators are defined in terms of the above distribution functions as follows:
\begin{enumerate}
	\item the mutual information:
\begin{equation}
\label{eq:I_def}
\fl \mathcal{I}_{\A:\B}  =
\sum_{{\sig}^{\A}}\sum_{{\sig}^{\B}}P({\sig}^{\A},{\sig}^{\B})\ln\left[\frac{P({\sig}^{\A},{\sig}^{\B})}{P_\M^\A({\sig}^{\A})P_\M^\B({\sig}^{\B})}\right],
\end{equation}
	\item and the separation entropy:
\begin{equation}
\label{eq:S_def}
 S_{\A:\B}  =  \sum_{{\sig}^{\A}}P({\sig}^{\A})\ln
P({\sig}^{\A})+\sum_{{\sig}^{\B}}P({\sig}^{\B})\ln P({\sig}^{\B}) -
\sum_{{\sig}^{\A},{\sig}^{\B}}P({\sig}^{\A},{\sig}^{\B})\ln
P({\sig}^{\A},{\sig}^{\B}).
\end{equation}
\end{enumerate}
These estimators can be written in a more compact form using the Shannon entropy
$H[P(\sig)]=-\sum_{\sig}P(\sig)\ln P(\sig)$, as
\begin{eqnarray}
\label{eq:Idef2}
I_{\A:\B}=H[P_\M^\A(\sig^\A)]+H[P_\M^\B(\sig^\B)]-H[P(\sig^{\A},\sig^\B)],  \\\nonumber
\label{eq:Sdef2}
S_{\A:\B}=H[P(\sig^{\A},\sig^\B)]-H[P(\sig^\A)]-H[P(\sig^\B)].
\end{eqnarray}
Of the two, the mutual information has been studied more extensively in the context of quantum systems \cite{Eisert2010}.

\subsection{Mean-field models}
\label{sec:models_intro}
It is in general quite difficult to compute the above information estimators for a
classical many-body systems in two or higher dimensions, and one has to resort to numerical methods
\cite{Iaconis2013}. The analytical calculations are simpler in
one-dimension, but the absence of phase transitions in short-range interacting
systems makes the computation redundant. For this reason, we consider
models with mean-field interactions which are known to exhibit rich phase
diagrams even in one dimension, providing non-trivial examples for studying
shared information.

In order to identify generic properties we consider a classical spin model with both mean-field and short-range
interactions of a general type. The model is defined on a one-dimensional lattice of length $L$. In the case where there is only mean-field
interactions, the notion geometry is lost and $L$ denotes simply the number of spins in the systems.
Every site is occupied by a $(p+1)$-state spin variable. We choose $p+1$ states rather than simply $p$ in order
to simplify the notation in the detailed calculation. We consider a general form of the mean-field interaction among the spins, defined
in terms of $Q_{k}$ variables with $k=1,\dots,p+1$, which counts the number of spins in
the $k^{\mathrm{th}}$ state and defined as
\begin{equation}
\label{eq:app_Qj}
Q_k(\sig)=\sum_{i=1}^{L} \delta_{l_k,\sigma_i}.
\end{equation}
Here $\delta$ denotes the Kronecker delta, $l_k$ is the value of $\sigma_i$ in the $k^{\mathrm{th}}$ state and $\sig$ denotes a spin
configuration of the entire system. We define the Hamiltonian as
\begin{equation}
\label{eq:Ham_generic}
\HH(\sig,L ) =  L ~\epsilon\Big(\frac{{\bf Q}(\sig)}{L}\Big)+\sum_{i,j} \phi_{i,j}(\sig),
\end{equation}
where ${\bf Q}=\{Q_1,\dots,Q_{p}\}$, and $\epsilon$ is an arbitrary function
which accounts for the mean-field interaction. The function $\epsilon$ depends only on $p$ parameters, rather than $p+1$, because the sum of the $Q$'s is always
$L$, i.e. $Q_{p+1}=L-\sum_{i=1}^p Q_i$.
 The second interaction term in \eref{eq:Ham_generic}, $\phi_{i,j}$,
represents a general short-range interaction potential among the spins, which
vanishes when $\vert i-j\vert$ is larger than some finite distance, $R$, which does not scale with
the system length \footnote{ For simplicity we focus here on the case of interval boundary conditions in the definition of $\phi_{i,j}$. The generalization of the derivation below to other boundary conditions is straightforward, and
their effect is found only in the constant term in the scaling form in \eref{eq:area_div}.}.
 Note that, because the ${\bf Q}(\sig)/L$ is intensive, the Hamiltonian remains extensive
in spite of the infinite-range interaction.

The above form of the Hamiltonian describes a large class of mean-filed models.
The two specific instances of the model which have been studied extensively in
the past are the mean-field Blume, Emery and Griffiths (BEG) model \cite{Blume1971,Barr2001} and the Nagel-Kardar (NK) model \cite{Nagle1970,Bonner1971,Kardar1983,Mukamel2005}.
Despite being one-dimensional, both models display a rich phase diagram. In the
following two subsections we present a brief description of the phase diagram that will be relevant for
our analysis.

\subsubsection{The mean-field BEG model:}
This is a three state spin model with variable
$\sigma_{i}=\left\{ -1,1,0 \right\}$ and a Hamiltonian
\begin{equation}
\label{eq:beg_ham}
\HH(\sig,L)=\Delta\sum_{i=1}^{L}\sigma_{i}^{2}-\frac{J}{2L}\left(\sum_{i=1}^{L}\sigma_{i}\right)^{2}.
\end{equation}
The parameter $\Delta$ is the on-site field strength and $J$ is the strength of the infinite range interaction between all the spins.
This is a special case of the general Hamiltonian in
\eref{eq:Ham_generic} with $p=2$ and $\epsilon(q_{1},q_2)=\Delta (q_1+q_2) -
J (q_2-q_{1})^2/2$ and $\phi_{i,j}(\sig)=0$.

\begin{figure}
\begin{center}
\includegraphics[scale=0.74]{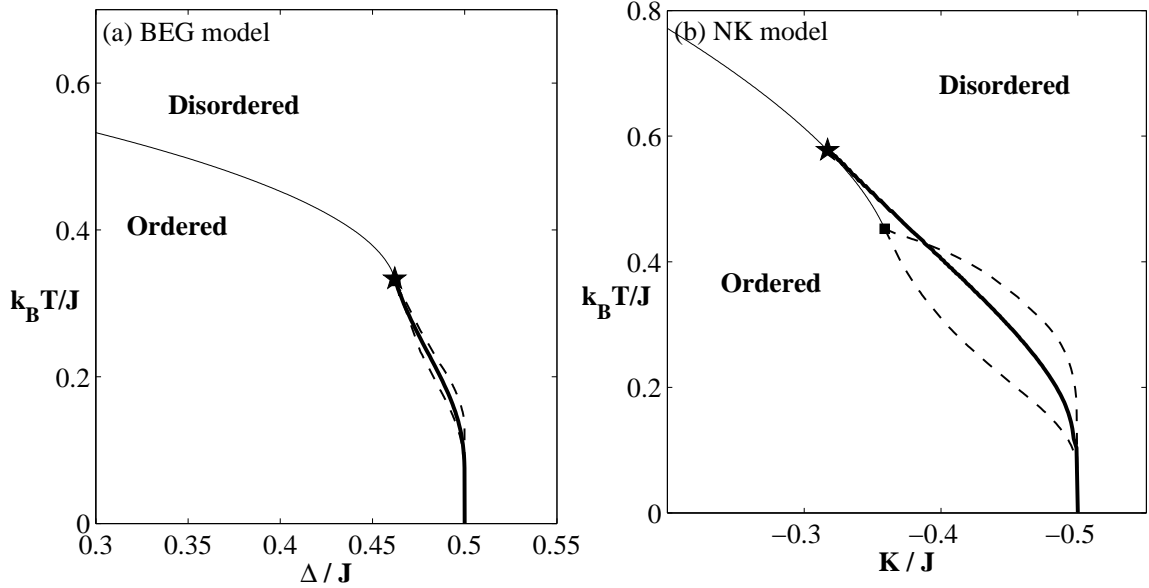}
\caption{Phase diagram of (a) the BEG model and (b) the NK model in the
canonical and the micro-canonical ensemble. For both models, the critical line
(thin, solid) separating the ordered and disordered phases, is identical
in the two ensembles. The line terminates at different tricritical points in each ensemble, denoted by a star ($\star$) in the canonical ensemble
and by a square ($\blacksquare$) in the micro-canonical ensemble. Below the
tricritical points the phases are separated by first order transition line,
denoted by thick solid line in the canonical ensemble and by dashed lines in the
micro-canonical ensemble. In the latter the intermediate region between the
dashed lines does not have a well defined temperature. In the BEG model, the two tricritical points are found to be
very close to each other, and thus appear to coincide in (a).
\label{fig:BEG}}
\end{center}
\end{figure}

The BEG model has been used in the past as a prototypical model of
long-range interacting systems, particularly in the study of ensemble inequivalence,
whereby a model exhibits different phase diagrams within two different ensembles \cite{Barr2001,Leyvraz2002,Ellis2004}. In both the micro-canonical and the canonical ensembles
the BEG model undergoes a phase transition between a paramagnetic (disordered) phase where
the average magnetization $ m = L^{-1} \langle \sum_{i}\sigma_i \rangle=0$, to
a ferromagnetic (ordered) phase, where $m\neq0$. The phase diagram in the two ensembles is shown
in \fref{fig:BEG}a. The temperature in the micro-canonical ensemble is defined by
the thermodynamic relation $T^{-1}=k_{B}\beta =\partial S/\partial E$ with
$S$ and $E$ being the entropy and the energy, respectively, and $k_B$ denoting the Boltzmann constant.
The two thermodynamic phases are separated by the following critical line:
\begin{equation}
\label{eq:critical line BEG}
\beta J=\frac{1}{2}\exp\left(\beta\Delta\right)+1,
\end{equation}
which meets a first order transition line at a tricritical point.

The inequivalence of the two ensembles can be seen in the position of the tricritical point and in the first order transition line.
In the canonical ensemble the first order transition is denoted by a thick solid line, where the average magnetization in the system,
$m$, changes discontinuously. In the micro-canonical ensemble the first order transition is denoted by two stability lines, which encompass a region
where the ordered and disordered phases are both either stable or meta-stable.
This inequivalence is a common feature in the long-range interacting systems \cite{Dauxois2009,Dauxois2010}.

\subsubsection{The NK model:}
This is a variant of the nearest-neighbor Ising spin chain with an additional
mean-field interaction term. The spin variables are $\sigma_{i}=\{-1,1\}$ and the Hamiltonian is given by
\begin{equation}
\label{eq:NK model_energy_first}
\HH(\sig,L)=-\frac{K}{2}\sum_{i=1}^{L-1}\left(\sigma_{i}\sigma_{i+1}-1\right)-\frac{J}{2L}\left(\sum_{i=1}^{L}\sigma_{i}\right)^{2}.
\end{equation}
The parameters $K$ and $J$ denote the short-range and long-range interaction strengths, respectively.
Similarly to the BEG model, the mean-field interaction strength is rescaled with
the system length $L$ to keep the energy extensive. This is a special case of the
general Hamiltonian \eref{eq:Ham_generic} with $p=1$, $\epsilon(q_1)= -\frac{J}{2}(2q_1-1)^2$ and $\phi_{i,j}(\sig) = - \frac{K}{2} (\sigma_{i}\sigma_{j}-1) \delta_{j-i,1}$.

The model has been studied within both the micro-canonical \cite{Nagle1970,Bonner1971,Kardar1983} and then canonical \cite{Mukamel2005}
ensembles. The phase diagrams corresponding to the two ensembles are shown in \fref{fig:BEG}b.
Similarly to the BEG model, the NK model exhibits a disordered phase with
vanishing average magnetization, and an ordered phase where the magnetization is
non-zero. At small values of the ratio $K/J$ the two phases are separated by a
second order transition line which in both the ensembles is given by
\begin{equation}
\beta J =\exp(-\beta K).
\end{equation}
As $K/J$ increases the second order transition line turns into a first order
line at a tricritical point, which is different for the two ensembles.
This ensemble inequivalence is qualitatively similar
to the one observed in the BEG model, as evident by the similarities between  \fref{fig:BEG}a and \fref{fig:BEG}b.

\section{The main results}
\label{sec:main}

In this section, we summarize the main results of our study of the information
estimators in the general model defined in
\eref{eq:Ham_generic}. A detailed derivation of these results is given in \sref{sec:generic_model}.
We consider the fictitious partitioning of the system into two subsystems, $\A$ and $\B$ of size $\ell$ and $L-\ell$, respectively, within two limits:
one where the sizes of both subsystems scales linearly with $L$, i.e. $1 \ll \ell \sim L$, and the other where
 $\ell$ is large but does not scale with $L$, i.e. $1\ll \ell \ll L$. Both of these limits
 have been considered in the past studies of entanglement in quantum systems. One would expect that the scaling
behavior of entanglement estimators in the second limit can be obtained by taking $\ell/L\to 0$ in the results obtained from the first limit ($1\ll \ell \ll L$).
 To our surprise, we
find that for the mutual information estimator at criticality this is not true. This difference between the two limits can be understood by a careful analysis, presented in \sref{sec:beg_finite}, which we also verify using a numerical computation.

The results below are presented first in the $\ell \sim L$ limit for the
canonical and microcanonical ensemble in
\sref{sec:sum_can} and \sref{sec:sum_mc}, respectively. The differences found in the $1\ll \ell \ll L$ limit are summerized in \sref{sec:sum_finite_size}.

\subsection{Canonical ensemble:}
\label{sec:sum_can}
{\it Away from the critical line}, both the
information estimators follow the area law, \textit{i.e.}, they remain finite as $L\to \infty$.
To leading order in $L$ we obtain that
\begin{eqnarray}
\label{eq:sum_I_beg_can_away}
I_{\A:\B} &=&\frac{1}{2}\ln\left[ g(\alpha)g(1-\alpha)\right] + \mathcal{O}(1),  \\
S_{\A:\B}&=& O(1),\nonumber
\end{eqnarray}
where $\alpha \equiv \ell / L$ is the fractional volume of the subsystem
$\A$ and $g$ is a scaling function.
The symbol $\mathcal{O}(1)$ denotes terms that do not increase with either $L$ or
$\ell$, and do not depend of $\alpha$. These terms involve the microscopic details of the model.

In general, we find that the scaling function $g(\alpha)$ has a non-universal form that
depends on the details in the Hamiltonian \eref{eq:Ham_generic}. For a
$(p+1)$-state spin chain, it is a polynomial of degree $p$ defined as
\begin{equation}
\label{eq:sum_g}
g(\alpha)= b_{p} \alpha^{p} + b_{p-1} \alpha^{p-1}+\ldots + b_2 \alpha^2 + \alpha+b_0,
\end{equation}
where $b_i$ depend on the details of the model. Note that, the coefficient of the
linear term is $1$.

Along the critical line, both the estimators diverge with $L$ and resemble the
scaling seen in the entanglement estimators in quantum systems. The mutual
information is given {\it at criticality} by
\begin{eqnarray}
\label{eq:sum_I_beg_can}
 I_{\A:\B}=\frac{1}{4}\ln \left[L
g(\alpha)^{2}g(1-\alpha)^{2}\right] + \mathcal{O}\left( 1 \right).
\end{eqnarray}
At criticality $g(0)$ vanishes and thus $b_0=0$.
In the example of the NK model, where $p=1$, this leads to a simple form of the scaling
function $g(\alpha)=\alpha$. For the BEG model, although $p=2$, the scaling
function is also give by $g(\alpha)=\alpha$. This is because the $\alpha^{2}$ term is excluded due to
a particular symmetry of the Hamiltonian, discussed in \sref{sec:gen_can}.

\begin{figure}
\begin{center}
\includegraphics[scale=0.65]{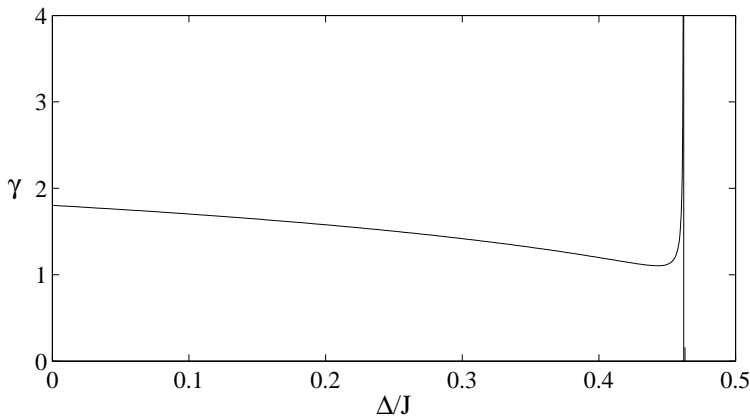}
\caption{\label{fig:gamma}
The coefficient $\gamma$ of the leading order term in the separation entropy at criticality in the canonical BEG model as a function of $\Delta/J$.
The coefficient diverges at the tricritical point, $\Delta/J \simeq  0.462$, where the critical line turns into a first order transition line. }
\end{center}
\end{figure}

For the separation entropy {\it at criticality} the leading divergence with $L$ is
$\sqrt{L}$ with a negative sub-leading logarithmic term. The overall scaling form is thus given by
\begin{eqnarray}
\label{eq:sum_S_beg_can}
\mathcal{S}_{\A:\B}=\gamma L^{1/2} \left[(1-\alpha)^{1/2}+\alpha^{1/2}-1\right]
-\frac{1}{4}\ln\left[ L \alpha(1-\alpha)\right]+\mathcal{O}(1),
\end{eqnarray}
where $\gamma$ is a non-universal coefficient whose explicit form is
derived in \sref{sec:beg_degeneracy}.
It is a strictly positive quantity resulting a positive separation entropy
$S_{\A:\B}$. We have verified this numerically for the BEG model as shown in the
\fref{fig:gamma}.
This is consistent with the fact that the entropy of the composite
system is higher than the combined entropy of the isolated subsystems.
Unlike the coefficient of the $\sqrt{L}$ term, the coefficient of the sub-leading
$\ln L$ term is universal and remains constant along the critical line.

\subsection{Micro-canonical ensemble:}
\label{sec:sum_mc}
In the micro-canonical ensemble the total energy of the system is strictly fixed. This global constraint is expected to induce
correlations between microscopic fluctuation in the two subsystems, and one
would expect the area law to break down. This is exactly what we find in
our analysis where both the information estimators have an additional
$\frac{1}{2}\ln L$ term  everywhere in the parameter space, even away from criticality.

{\it Away from criticality} we find the following leading $L$ dependence of the shared-information estimators:
\begin{eqnarray}
 I_{\A:\B}
&=&\frac{1}{2}\ln\left[ L g(\alpha)g(1-\alpha)\right] +
\mathcal{O}(1),\label{eq:I_mc_away}  \\
S_{\A:\B}&=&\frac{1}{2}\ln\left[ L \alpha(1-\alpha) \right]
+O(1),\label{eq:sum_mc_away}
\end{eqnarray}
whereas {\it at criticality} it changes into
\begin{eqnarray}
I_{\A:\B}
&=&\frac{3}{4}\ln \left[ L (g(\alpha)g(1-\alpha))^{2/3}\right]+
\mathcal{O}(1), \label{eq:Iab micro critical} \\
S_{\A:\B}&=&\frac{1}{4}\ln\left[ L \alpha(1-\alpha) \right]
 +\mathcal{O}(1).\label{eq:S_micro}
\end{eqnarray}

\begin{figure}
\begin{center}
\includegraphics[scale=0.65]{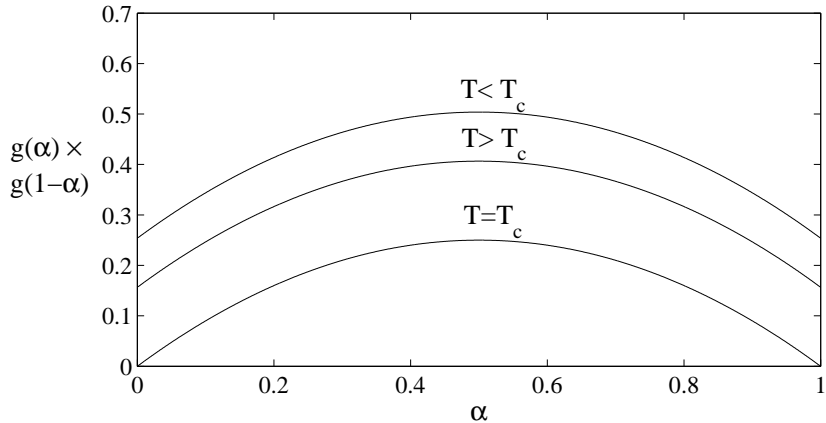}
\caption{\label{fig:scaling_I}
 The scaling function of the leading order term in the mutual information in the micro-canonical BEG model, $g(\alpha)g(1-\alpha)=(b_0+\alpha)(b_0+(1-\alpha))$.
 The function is plotted for $\Delta/J=0.35$ and for different values of the
 energy corresponding to the homogeneous, critical and ordered phases.
 For convenience we provide the corresponding values of the temperature defined
 in the micro-canonical ensemble using the relation $T=(ds/d\epsilon)^{-1}$.
 In the homogeneous phase $k_BT/J=0.6$ and $b_0\simeq 0.137$, in the critical phase $k_BT_c/J\simeq 0.497$ and $b_0=0$ and in the ordered phase $k_B T/J=0.49$ and
 $b_0\simeq 0.21$.}
\end{center}
\end{figure}

The scaling function $g(\alpha)$ has the same form as in \eref{eq:sum_g}, with
the coefficients $b_i$ depending on the microscopic details of the model.
For the BEG model, we find that the scaling function has a simple form
$g(\alpha)=\alpha+b_{0}$, where the constant $b_{0}$ vanishes along the critical
line. A plot of the scaling function $g(\alpha)g(1-\alpha)$ for the
BEG model is found in \fref{fig:scaling_I} for representative points in the parameter-space.

Another notable feature is the absence of the leading $\sqrt{L}$ term in
\eref{eq:S_micro} as compared to the form of the separation entropy  in the canonical ensemble, given in
\eref{eq:sum_S_beg_can}. This $\sqrt{L}$ divergence in the canonical ensemble results from finite-size corrections to the total energy, as discussed below \eref{eq:gamma1}. In the micro-canonical ensemble where the total energy is strictly fixed this term vanishes.

It is important to stress that the additional $\frac{1}{2}\ln L$ terms observed in the micro-canonical ensemble are due to the fixed energy constraint.
In general, such $\ln L$ terms are related to long-range correlations which in our case can result from either the explicit long-range interactions
or from the total energy constraint. The source of the $\frac{1}{2}\ln L$ term can be verified by setting
 the mean-field interaction term to zero in our model, resulting in a model with only short-range interactions.
Following the derivation presented below one obtains a similar $\frac{1}{2}\ln L$ difference between the canonical and microcanonical calculations,
which implies that this difference is indeed due to the total energy constraint.

\subsection{Small $\ell/L$ scaling:\label{sec:sum_finite_size}}

As mentioned above, in most cases studied here the scaling behavior of the information estimators in the limit $1\ll \ell \ll L$ can be obtained by
taking $\alpha\equiv \ell/L \to 0$ in the results presented above. The only exception is the behaviour of the mutual information estimator at criticality, for which the $\ell/L\to0$ and $L\to \infty$ limits do not commute. Before we discuss this case, let us summarize the results where this procedure does work.

In the canonical ensemble and {\it away from criticality}, since the scaling function
$g(\alpha)$ in \eref{eq:sum_g} generally obeys $g(0)\neq0$, the information estimators
\eref{eq:sum_I_beg_can_away} are finite when considering $\alpha\rightarrow 0$, i.e.
\begin{eqnarray}
 I_{\A:\B}=\mathcal{O}(1),
\qquad S_{\A:\B}=\mathcal{O}(1). 
\end{eqnarray}
The same reasoning works in the non-critical micro-canonical ensemble, where the
scaling forms in \eref{eq:I_mc_away}-\eref{eq:sum_mc_away} yield in the $\alpha\rightarrow 0$ limit
\begin{eqnarray}
  I_{\A:\B} = &\frac{1}{2}\ln L + \mathcal{O}(1),\qquad
  S_{\A:\B}= &\frac{1}{2}\ln \ell +O(1).
\end{eqnarray}

{\it At criticality} in the canonical ensemble taking $\alpha\rightarrow 0$ limit in \eref{eq:sum_S_beg_can} yields
\begin{equation}
S_{\A:\B}= \gamma \ell^{1/2} -\frac{1}{4}\ln \ell +\mathcal{O}(1),
\end{equation}
whereas in the micro-canonical ensemble taking the same limit in \eref{eq:S_micro} leads to
\begin{equation}
	S_{\A:\B}=\frac{1}{4}\ln \ell +\mathcal{O}(1).
\end{equation}

Taking the same $\alpha\rightarrow 0$ limit in the expression for mutual
information \eref{eq:sum_I_beg_can} at criticality leads to a negative, diverging result
since $g(0)=0$. This cannot be the right result as the mutual
information is a positive quantity. A detailed microscopic derivation in the
limit $1\ll \ell \ll L$, given in \sref{sec:beg_finite}, yields the correct result whereby
\begin{equation}
I_{\A:\B}=\mathcal{O}(1),
\end{equation}
in the canonical ensemble {\it at criticality}. A similar analysis within the micro-canonical ensemble
shows that the mutual information diverges {\it at criticality} with the system size $L$, as
\begin{equation}
I_{\A:\B} =\frac{1}{2}\ln L + \mathcal{O}(1).
\end{equation}
Similarly to the $\ell \sim L$ limit, this divergence can be attributed to the fixed total energy constraint.

\section{Explicit calculation of the information estimators}
\label{sec:generic_model}
In this section we compute the scaling form of the shared-information for large $L$ using the saddle point method. This calculation is relatively straightforward for pure mean-field models.
 As shown below, the inclusion of the short-range interactions
 in the generic Hamiltonian (\ref{eq:Ham_generic}) does not alter the derivation significantly and affects only the sub-leading terms in $L$ of the information estimators. The calculation
 is first carried out in detail within the canonical ensemble, in \sref{sec:gen_can}, where it is relatively simple. For the microcanonical case, we present a sketch of the calculation in \sref{sec:gen_micro}. Additional issues which include ground-state degeneracy, small $\ell$ scaling and higher order critical points are discussed in section \ref{sec:beg_degeneracy}, \ref{sec:beg_finite} and \ref{sec:higher_order}, respectively.

\subsection{Canonical ensemble}
\label{sec:gen_can}
We consider first the generic model within the canonical ensemble, where the probability of a micro-state of the whole system, $\sig$, is given by
 \begin{equation}
 P(\sig)= Z^{-1}e^{-\beta\HH(\sig,L)}.
 \end{equation}
 The partition function, $Z$, is defined as
\begin{equation}
\label{eq:Z_can_delta}
Z=\sum_{\sig}e^{-\beta\HH(\sig)}=\sum_{\sig}e^{-L\beta\epsilon\big({\bf Q}(\sig)/L\big)-\beta\sum_{i,j}\phi_{i,j}(\sig)}.
\end{equation}
The first step in computing the shared-information estimators is the computation of the partition function.

\subsubsection{Partition function:}

The computation of $Z$ can be carried out using a standard technique, employed repeatedly in this paper, by which
 we replace the $e^{-L\beta\epsilon\big({\bf Q}(\sig)/L\big)}$ with an integral over a continuous variable ${\bf q}$, yielding
\begin{equation}
e^{-L\beta\epsilon\big({\bf Q}(\sig)/L\big)} = \int d{\bf q}e^{-L\beta\epsilon({\bf q})}  \prod_{j=1}^{p}\delta\big( Q_{j}(\sig)/L- q_j\big),
\end{equation}
where $\delta$ denotes the Dirac delta function.  For the partition function, this procedure yields
\begin{equation}
Z=L^p \int d{\bf q} e^{-L\beta\epsilon({\bf q})}\sum_{\sig}e^{-\beta\sum_{i,j}\phi_{i,j}(\sig)}\prod_{j=1}^{p}\delta\big( Q_{j}(\sig)-L q_j\big).
\end{equation}
The delta function can be replaced by a set of integrals over a $p$-vector-field ${\bf h}$, yielding
\begin{equation}
\label{eq:Z_can_0}
Z= L^p\int d{\bf q} d {\bf h} e^{-L\beta[\epsilon({\bf q})+\mathbf{h}\cdot\mathbf{q}]}\sum_ {\sig}e^{-\beta\sum\limits_{i,j}\phi_{i,j}(\sig)+\beta\mathbf{h}\cdot\mathbf{Q}(\sig)}.
\end{equation}

The sum over $\sig$ in \eref{eq:Z_can_0} is in fact the partition sum of
 a system with a short-range interaction term,
$\phi_{i,j}(\sig)$, and a field conjugate to $Q_{j}(\sig)$, denoted by ${\bf h}$, whose Hamiltonian is thus given by
\begin{equation}
\label{eq:Ham_short}
\HH'(\sig)=\sum_{i,j}\phi_{i,j}(\sig)-{\bf h}\cdot {\bf Q}(\sig).
 \end{equation}
In the following we demonstrate that the partition sum of this Hamiltonian is given to leading order in $L$ by $e^{L\ln\lambda_1+\mathcal{O}(1)}$,
where $\lambda_1$ is the largest eigenvalue of the transfer matrix corresponding to $\HH'$, and $\mathcal{O}(1)$ denotes terms that do not scale with $L$.
The crucial point is that there are no $\ln L$ terms in the exponent.

To this end, the partition sum is written in terms of transfer matrices as
\begin{equation}
\label{eq:Z_can_1}
\sum_ {\sig}e^{-\beta\sum_{i,j}\phi_{i,j}(\sig)+\beta\mathbf{h}\cdot\mathbf{Q}(\sig)}=\langle 1 | T_{\beta,\mathbf{h}}^{L}|1\rangle
\end{equation}
where $T_{\beta,\mathbf{h}}$ is the transfer matrix corresponding to the $\HH'$ and $\langle 1 |, |1 \rangle$ are the left and right identity vectors.

An important property of $T_{\beta,\mathbf{h}}$ is that its dimension does not scale with $L$.
In the case $\phi_{i,j}$ describes only nearest neighbours interactions, denoted by $\phi_{i,j}(\sig)=K_{\sigma_i, \sigma_{j}}\delta_{j,i+1}$, the transfer matrix is of dimension $(p+1)\times (p+1)$ and it is given by
\begin{equation}
 \big(T_{\beta,\mathbf{h}}\big)_{q,r}=e^{-\beta (K_{q,r}-h_q)}.
\end{equation}
For more general interaction range, $R$, the transfer matrix is constructed similarly, but by taking into account the state of the $R$ nearest neighbours. The dimension of the matrix is therefore at most $(p+1)^R\times (p+1)^R$. The fact that $T_{\beta,\mathbf{h}}$ is of finite dimension implies by the Perron-Frobenius theorem that its largest and its second largest eigenvalues
differ by a gap which is independent of $L$. Denoting the eigenvalues by $\lambda_k$ and the corresponding eigenvectors by $|v_k\rangle$, we obtain that
\begin{eqnarray}
\label{eq:Z_can_2}
\langle 1|T_{\beta,\mathbf{h}}^{L}|1\rangle=\lambda_1^L\big( \langle 1|v_1\rangle\big)^2 \big[1+\mathcal{O}(|\lambda_2|^L/\lambda_1^L)\big].
\end{eqnarray}

Inserting the leading order term in \eref{eq:Z_can_2} into the partition sum in \eref{eq:Z_can_0}, one obtains an integral which can be evaluated
in the $L\to \infty$ limit using the saddle point approximation (SPA) \footnote{
 The equivalence of statistical ensembles in short-range interacting systems implies that $\lambda_1$ is a convex function of ${\bf h}$. This allows us to perform
 the Laplace transform  \eref{eq:Z_can_delta} and the corresponding inverse transform to obtain the correct leading order contribution to the original sum. }.
We denote the result of the approximation of the integral over ${\bf h}$ by
\begin{equation}
\label{eq:Z_can_3}
\int d{\bf h} e^{-L\beta\mathbf{h}\cdot\mathbf{q}}\left\langle 1| T_{\beta,\mathbf{h}}^{L}|1\right\rangle = \omega(\mathbf{q}) e^{-L\beta f_\phi(\beta,{\bf q})}\big[1+\mathcal{O}(L^{-1})\big] ,
\end{equation}
where $f_\phi(\beta,{\bf q})$ denotes the Landau free energy density of ${\bf q}$ in a system with only short-range interactions, given by $\phi_{i,j}(\sig)$.
The function $\omega(\mathbf{q})$ accounts for the $\mathcal{O}(1)$ pre-factor of the leading order term in $\langle 1| T_{\beta,\mathbf{h}}^{L}|1\rangle$ and additional
pre-factors that result from the saddle point approximation. A specific example of $f_\phi(\beta,{\bf q})$ and $\omega({\bf q})$ can be obtained
 for pure mean-field models, where $\phi_{i,j}(\sig)=0$. In this case one obtains from combinatorial considerations that
\begin{eqnarray}
f_\phi(\beta,{\bf q}) &=& \frac{1}{\beta}\sum_{j=1}^p q_j \ln q_j - \frac{1}{\beta}(1-\sum_{j=1}^p q_j) \ln (1-\sum_{j=1}^p q_j) \equiv - \frac{1}{\beta} s_0({\bf q}) ,\\
\omega({\bf q}) &=&  \Big[ \pi^p (\prod_{j=1}^p q_j) (1-\sum_{j=1}^p q_j)\Big]^{-1/2} \equiv \omega_0({\bf q}).
\end{eqnarray}
Here $s_0({\bf q})$ is the entropy of a noninteracting spin system for a given value of ${\bf q}$.
Inserting \eref{eq:Z_can_3} into the partition function in \eref{eq:Z_can_1} yields finally
\begin{equation}
  \label{eq:app_Z2}
 Z=L^{p/2}\int d{\bf q} \omega(\mathbf{q}) e^{-L\beta f_\HH(\beta,{\bf q})}\big[1+\mathcal{O}(L^{-1})\big],
\end{equation}
where $f_\HH(\beta,{\bf q})=\epsilon({\bf q})+f_\phi(\beta,{\bf q})$ is the Landau free energy density of the complete system, which includes
the short-range and the long-range interaction terms.

 This integral in \eref{eq:app_Z2} can be further evaluated using the SPA, which we choose to separate into two steps. In the first step we approximate the integrals over
 $q_2,\dots,q_p$, by expanding the exponent to quadratic order in these variables. The next step is to approximate the remaining one-dimensional
 integral, by expanding the exponent to order $q_1^2$ away from criticality and to order $q_1^4$ at criticality.
 In cases where $p=1$, such as in the NK model, the first step is skipped.
 The first step of the SPA yields,
 \begin{equation}
 \label{eq:app_Z_first_step}
   Z =  L^{1/2} \int dq_1 \tilde{\omega}(q_1) e^{-L \beta f_\HH(\beta,\tilde{{\bf q}})}\big[1+\mathcal{O}(L^{-1})\big]
  \end{equation}
 where $\tilde{\bf q}(q_1)=\left(q_1,{\tilde q}_2(q_1),{\tilde q}_3(q_1),\dots, {\tilde q}_p(q_1)\right)$ is the solution of the set of equations $\partial f_\HH/ \partial q_i =0$
 for $i=2,3,\dots,p$.
The function $\tilde{\omega}(q_1)$ accounts for the contribution from $\omega (\tilde{{\bf q}})$ and the coefficients that results from the SPA,
  \begin{equation}
\label{eq:omega_tilde}
   \tilde{\omega}(q_1)\equiv \omega (\tilde{{\bf q}}(q_1)) \pi^{(p-1)/2} \Big[ \det \Big( \frac{\beta \partial^2 f_\HH}{\partial q_i \partial q_j} \Big\vert_{\tilde{{\bf q}}(q_1)} \Big)\Big]^{-1/2},
  \end{equation}
 for $i,j=2,3,\dots,p$.

{\it Away from the critical line}, the SPA over of the integral over  $q_1$ in \eref{eq:app_Z_first_step}  yields
\begin{eqnarray}
 \label{eq:app_Z_away}
 Z  \simeq \tilde{\omega}(q^\star_{1}) e^{-L \beta f_\HH(\beta,{\bf q}^\star)}  \chi_{0,2}\left( \beta \frac{d^2 f_\HH(\tilde{{\bf q}}(q_1))}{d q_1^2} \Big\vert_{q^\star_{1}} \right),
\end{eqnarray}
where ${\bf q}^\star = (q^\star_{1},q^\star_{2},\dots,q^\star_{p})$ is the global minimum of $f_\HH$ and $\chi_{2,0}(a)=\sqrt{\pi/a}$ denotes the coefficient
which results from the Gaussian integral.
 Since such integrals are performed  frequently in the rest of the paper, it is convenient to define
the following notation:
\begin{equation}
\label{eq:general_gausian_integral}
\fl\qquad
\int_{-\infty}^{\infty}dx x^{r}e^{-L a
x^{s}}=\frac{1}{L^{\frac{r+1}{s}}}\left[\frac{2}{s\times
a^{\frac{r+1}{s}}}\Gamma(\frac{r+1}{s})\right] \equiv \frac{1}{L^{\frac{r+1}{s}}}\chi_{r,s}(a),
\end{equation}
where $a>0$ and $r,s$ are positive even integers. In this section and in \sref{sec:gen_micro} we assume that the Landau free energy, $f_\HH(\beta,{\bf q})$, has a single global minimum. The effect
of degenerate minima is discussed in \sref{sec:beg_degeneracy}, where the degeneracy is shown to affect only the $\mathcal{O}(1)$ term of the information estimators.

On the critical line, we assume without loss of generality that the determinant in \eref{eq:omega_tilde} does not vanish.
 This implies that the order parameter of the transition is a combination of the $q$'s that necessarily involves $q_1$. On the other hand, the argument
  of $\chi_{0,2}$ in \eref{eq:app_Z_away} does vanish at criticality. As a result, the SPA of the integral  in \eref{eq:app_Z_first_step} has to be carried out by expanding the
exponent in \eref{eq:app_Z_first_step} to order $q_1^4$, yielding {\it at criticality} the following scaling form:
\begin{eqnarray}
 \label{eq:app_Z_at}
 Z  \simeq L^{1/4} \tilde{\omega}(q^\star_{1}) e^{-L\beta f_\HH(\beta,{\bf q}^\star)}  \chi_{0,4}\left( \beta \frac{d^4f_\HH(\beta,\tilde{{\bf q}}(q_1))}{d q_1^4} \Big\vert_{q^\star_{1}}\right).
\end{eqnarray}
The resulting expressions for $Z$ will be used below in the derivation of $S_{\A:\B}$ and $\mathcal{I}_{\A:\B}$.

\subsubsection{Separation entropy:}
The Shannon separation entropy, $S_{\A:\B}$, can be derived directly from the expression Shannon entropy of the whole system, given by
\begin{equation}
\label{eq:S3}
 S=-\sum_{\sig} P(\sig) \ln P(\sig) = \ln Z+Z^{-1}\sum_{\sig}e^{-\beta\HH(\sig)}\beta\HH(\sig).
\end{equation}
Following a derivation similar to that of $Z$, the entropy can be expressed in terms of the transfer matrix of a short-range interacting system, whose Hamiltonian is $\HH'$, as
\begin{equation}
 S=\ln Z+Z^{-1}L^p\int d{\bf q} d{\bf h} e^{-L\beta[\epsilon({\bf q})+\mathbf{h}\cdot\mathbf{q}]}\Big[L\beta\epsilon({\bf q})\langle 1| T_{\beta,\mathbf{h}}^{L}|1\rangle -\frac{\partial}{\partial\beta}\langle 1| T_{\beta,\mathbf{h}}^{L}|1\rangle\Big].
\end{equation}
Similarly to the derivation of \eref{eq:app_Z2}, in the $L\to\infty$ limit one may consider only the leading order term of $\langle 1| T_{\beta,\mathbf{h}}^{L}|1\rangle$
and evaluate the integral over ${\bf h}$ of  using the SPA, yielding
\begin{equation}
\label{eq:S_int}
 S=\ln Z+Z^{-1}L^{p/2}\int d{\bf q}\omega(\mathbf{q})e^{-L\beta f_\HH(\beta,\mathbf{q})}L[\beta\epsilon(\mathbf{q})+\beta\phi(\beta,\mathbf{q})]+\mathcal{O}(1),
\end{equation}
where $\phi(\beta,\mathbf{q})\equiv\frac{\partial}{\partial \beta} [\beta f_\phi(\beta,\mathbf{q})]$ is the average energy of the short-range interacting
system, when constrained on a specific value of the coarse variables, ${\bf Q}(\sig)=L\mathbf{q}$.
One can define in a similar manner the average entropy of this system, $s({\bf q})\equiv-\beta f_\phi(\beta,\mathbf{q})+\beta\phi(\beta,\mathbf{q})$, which will be used below.

As in the case of $Z$, \eref{eq:S_int} can be evaluated using a two-step saddle point approximation of the integral over ${\bf q}$. In the first step, the SPA of the integrals over $q_2,\dots,q_p$ yields
\begin{equation}
 \label{eq:app_S3}
S  \simeq -\ln Z - L \beta\frac{  \int d q_1 \tilde{\omega}(q_1) [ \epsilon(\tilde{{\bf q}}(q_1))+\phi(\beta,\tilde{{\bf q}}(q_1))] e^{-L \beta f_{\HH}(\beta,\tilde{{\bf q}}(q_1))} }
{\int d q_1 \tilde{\omega}(q_1)e^{-L \beta f_{\HH}(\beta,\tilde{{\bf q}}(q_1))} }.
\end{equation}
 The SPA of the integral over $q_1$ is done by expanding the exponent to order $q_1^2$ {\it away from criticality} yielding
\begin{equation}
 \label{eq:app_S_away}
S  = L  s({\bf q}^\star) + \mathcal{O}(1).
\end{equation}
On the other hand, {\it at criticality} the exponent need to be expanded to order $q_1^4$, leading to the following scaling form:
\begin{equation}
 \label{eq:app_S_at}
S  = L s({\bf q}^\star) + \gamma L^{1/2} - \frac{1}{4} \ln L + \mathcal{O}(1).
\end{equation}
Note that in deriving \eref{eq:app_S_away} and \eref{eq:app_S_at} $Z$ has been replaced by its expression in \eref{eq:app_Z_away} and \eref{eq:app_Z_at}, respectively.
The coefficient $\gamma$, given by
\begin{equation}
\label{eq:gamma1}
\gamma = \beta \frac{\chi_{2,4}(\beta\frac{d^4 f_{\HH}}{d q^4_1}|_{q_1^\star})}{\chi_{0,4} (\beta\frac{d^4 f_{\HH}}{d q^4_1}|_{q_1^\star}) \omega({\bf q}^\star)} \frac{d}{dq_1^2} \big\{ \tilde{\omega}(q_1)[\epsilon( \tilde{\bf q}(q_1))+\phi(\beta, \tilde{\bf q}(q_1)) - \epsilon({\bf q}^\star)-\phi(\beta,{\bf q}^\star) ] \big\} \Big|_{q_1^\star},
 \end{equation}
 depends in general on the parameters of the model. Its form suggests that it is related to the finite-size corrections to the mean energy.
This can be clearly understood by noting that the source of the $\gamma L^{1/2}$ term is the second term in the RHS of \eref{eq:S3}, which corresponds to the average energy in the systems, i.e. $Z^{-1}\sum_{\sig}e^{-\beta\HH(\sig)}\beta\HH(\sig) = \langle \HH \rangle = [\epsilon({\bf q}^\star)+\phi(\beta,{\bf q}^\star)]  L + \gamma L^{1/2} + O(\log L)$.
 This coefficient is studied in more detail in the BEG model in \fref{fig:gamma}, where it is found shown to be
strictly positive.
This coefficient was found to diverge at the tricritical point, where
$\frac{d^2 f_{\HH}}{d q^2_1}|_{q_1^\star}=\frac{d^4 f_{\HH}}{d q^4_1}|_{q_1^\star} =0$, indicating that the exponent in \eref{eq:app_S3} has to be expanded to order $q_1^6$ in
order to obtain the correct scaling of $S$. This divergence is evident in \fref{fig:gamma}. The behaviour of $S$ at tricritical points is discussed in \sref{sec:higher_order}.

\begin{figure}
\begin{center}
\includegraphics[scale=0.6]{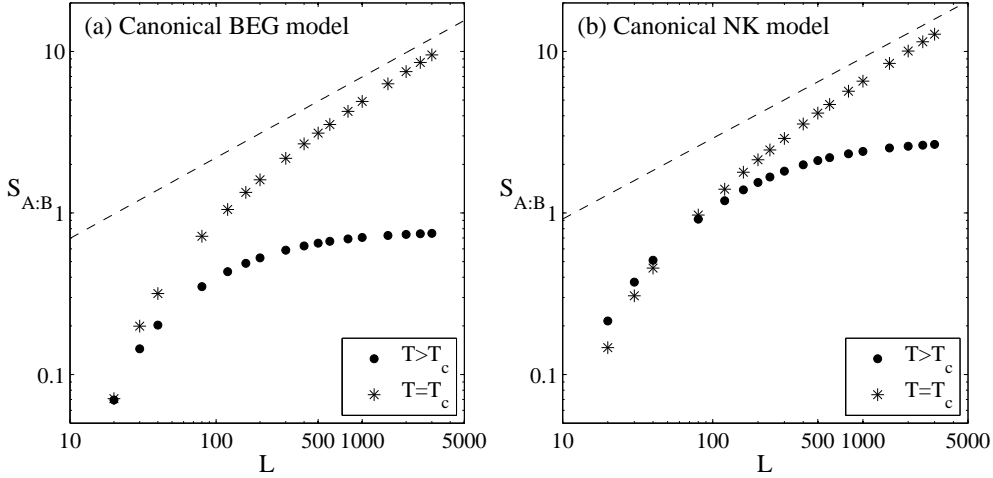}
\caption{
Log-log plot of the Shannon separation estimator in the canonical ensemble as a function of the system-size for $\ell/L=1/2$.
Figure (a) shows the results for the BEG model at criticality ($\Delta/J\approx 0.379$, $k_B T/J\approx 0.473$) and away from criticality ($\Delta/J\approx 0.441$, $k_B T/J\approx 0.552$), denoted by $\star$ and $\bullet$ respectively.
Similarly, figure (b) shows the results for the NK model  at criticality ($K/J\approx -0.268$, $k_B T/J\approx 0.670$) and away from criticality ($K/J\approx -0.287$, $k_B T/J\approx 0.718$).
At criticality, the leading order term in $S_{\A:\B}$ is expected to scale as $\sqrt{L}$, depicted by the dashed lines, whereas away from criticality $S_{\A:\B}$ is expected to converge to a constant.
\label{fig:can_SAB}}
\end{center}
\end{figure}

As mentioned above, the separation entropy measures the difference between the Shannon entropy
of the whole system and that of the two subsystems when they are physically decoupled.
The two decoupled subsystems $\A$ and $\B$, are assumed to obey the Gibbs-Boltzmann distribution
with respect to the Hamiltonian, $\HH(\sig,\ell )$ and $\HH(\sig,L-\ell )$, respectively, where $\HH$ is defined in \eref{eq:Ham_generic}. This implies
that after the separation the interaction strength in the Hamiltonian of each subsystem
has to be rescaled with the size of the each subsystem. This rescaling ensures that the separated subsystems would have the same
 values of ${\bf q}^\star$ as those of the composite system.
Since the decoupled subsystems maintain the form of the Hamiltonian of the whole system, their Shannon entropies are given by the above expression with the size
$L$ replaced by $\ell$ for subsystem $\A$ and by $L-\ell$ for subsystem $\B$.
As a result the extensive terms in the separation entropy in \eref{eq:S_def} cancel and we obtain that {\it away from criticality} to leading order

\begin{eqnarray}
\label{eq:can_S_away}
S_{\A:\B}=\mathcal{O}\left( 1 \right).
\end{eqnarray}
The cancelation of the $\mathcal{O}(L)$ terms in $S_{\A:\B}$ suggests that the rescaling of the Hamiltonians, described above, is a physically sensible way to define the separation process.
{\it At criticality}, the extensive terms still cancel but the $\sqrt{L}$ and $\log L$ terms do not, yielding
\begin{eqnarray}
\label{eq:can_S_at}
S_{\A:\B} = \gamma L^{1/2}\left[\sqrt{\alpha}+\sqrt{1-\alpha}- 1
\right]-\frac{1}{4}\ln\left[ L \alpha \left( 1-\alpha \right)
\right]+\mathcal{O}\left( 1 \right).
\end{eqnarray}
The results in \eref{eq:can_S_away} and \eref{eq:can_S_at} are verified numerically in \fref{fig:can_SAB} for the BEG and NK models.

\subsubsection{Mutual information:}
The mutual information estimator is based on the marginal probability distribution of the bipartition, which for subsystem $\A$ is defined as
$P_{\M}^{\A}({\sig}^{\A})=\sum_{{\sig}^{\B}}P({\sig}^{\A},{\sig}^{\B})$. The derivation below is done mainly for subsystem $\A$. The results for subsystem $\B$
can be obtained by replacing $\A \to \B$ and $\ell \to L-\ell$ in the expressions below. The marginal distribution of $\A$ can be written as
\begin{equation}
\label{eq:PMA_three_term}
P_{\M}^{\A}({\sig}^{\A})= \frac{1}{Z}e^{-\beta\sum_{i,j\in\A}\phi_{i,j}(\sig^\A)} Z^{\B}(\mathbf{Q}({\sig}^{\A})/L,\sig^\A)
\end{equation}
where $Z^{\B}(\mathbf{q}^{\A},\sig^\A)$ is the partition function of subsystem $\B$, defined as
\begin{equation}
Z^{\B}(\mathbf{q}^{\A}, \sig^\A) \equiv \sum_{{\sig}^{\B}}e^{-L\beta \epsilon(\mathbf{q}^\A+\mathbf{Q}({\sig}^{\B})/L)-\beta\sum_{i\in \B,j \in \B}\phi_{i,j}(\sig)-2\beta\sum_{i\in \A,j \in \B}\phi_{i,j}(\sig)}.
\end{equation}
Note that the term $\sum_{i\in \A, j\in \B}\phi_{i,j}(\sig)$ in the exponentials corresponds to the short-range interactions on the boundary between the two subsystems,
and thus involves a number of terms that does not scale with $L$. This fact will be used below to neglect its contribution.

In the $L\to\infty$ limit the leading order term in $Z^\B$ can be simplified using the same technique employed in the computation of $Z$ above \eref{eq:app_Z2}, yielding
\begin{eqnarray}
\label{eq:ZBa}
\fl \qquad \qquad Z^{\B}(\mathbf{q}^{\A},\sig^\A)&=&
(L-\ell)^{p}\int d{\bf q}^{\B} d{\bf h}^{\B}
e^{-L\beta[\epsilon(\mathbf{q}^{\A},\mathbf{q}^{\B})+{\bf h}^\B \cdot \mathbf{q}^{\B}] }\big\langle \sig^{\A}_{\mathrm{b}}|T_{\beta,\mathbf{h}^{\B}}^{L-\ell}|1\big\rangle  \\
\fl \qquad \qquad &=&(L-\ell)^{p/2}\int d{\bf q}^{\B}e^{-(L-\ell)\beta f_{\HH,1-\alpha}(\beta,{\bf q}^\A,{\bf q}^\B)}\omega(\mathbf{q}^{\B},\sig^\A)
\big[1+\mathcal{O}(L^{-1})\big], \nonumber
\end{eqnarray}
where
\begin{equation}
\label{eq:app_f_def}
f_{\HH,x}(\beta,{\bf q}^\A,{\bf q}^\B) \equiv \frac{1}{\beta x}\epsilon( (1-x) \mathbf{q}^{\A}+ x \mathbf{q}^{\B})+f_\phi(\beta,\mathbf{q}^{\B}).
\end{equation}
The function $f_{\HH,x}(\beta,{\bf q}^\A,{\bf q}^\B)$ can be regarded as the Landau free energy corresponding of a single subsystem of size $xL$, given the values of the coarse variables in the complementary subsystem, denoted by ${\bf q}^\A$. The vector $\big\langle \sig^{\A}_{\mathrm{b}}|$ in \eref{eq:ZBa} denotes configuration of the {\it boundary spins} in
subsystem $\A$ which interact with subsystem $\B$.
This boundary condition affects only the $\mathcal{O}(1)$ coefficient of the leading order term in $L$, denoted by $\omega(\mathbf{q}^{\B},\sig^\A)$.

In general, evaluating the integral in \eref{eq:ZBa} involves a complicated expression for the saddle point of the integrand as a function of ${\bf q}^\A$.
However as will be shown below, for the purpose of computing the $\ln L$ terms in the mutual information it is sufficient to evaluate $Z^\B$ only for ${\bf q}^\A={\bf q}^\star$,
where ${\bf q}^\star$ is the saddle point of Landau free energy of the full system, $f_{\HH}(\beta,{\bf q})$.
For the case that ${\bf q}^\A={\bf q}^\star$, it can be easily shown that the saddle point of $f_{\HH,1-\alpha}(\beta,{\bf q}^\star,{\bf q}^\B)$
is found at ${\bf q}^\B={\bf q}^\star$ as well.
In contrast to the calculation the partition function, the quadratic terms in $q_j^\B$ of the exponential in \eref{eq:ZBa} do not vanish at criticality (since $f_{\HH,x}(\beta,{\bf q}^\star,{\bf q})\neq
f_\HH(\beta,{\bf q})$ for $x<1$). This implies that the SPA of the integral in \eref{eq:ZBa} yields the same results both {\it at criticality} and {\it away from criticality}, given by
\begin{equation}
\label{eq:app_PMA}
Z^\B ( {\bf q}^\star,\sig^\A) = \omega({\bf q}^\star,\sig^\A) \pi^{p/2} \left[\det( A_{1-\alpha})\right]^{-1/2} e^{-L(1-\alpha)\beta f_{\HH,1-\alpha}(\beta,{\bf q}^\star,{\bf q}^\star)}\big[1+\mathcal{O}(L^{-1})\big],
\end{equation}
where the Hessian matrix $A_{1-\alpha}$ is given for a general subsystem-size by
\begin{equation}
\label{eq:app_A_def}
(A_{x})_{i,j}= \beta \frac{\partial ^2  f_{\HH,x} }{\partial q_i^\B \partial q_j^\B} \Big\vert_{{\bf q}^\B={\bf q}^\star}.
\end{equation}

The Shannon entropy of the marginal probability distribution, denoted here by $S_{\M}^{\A}\equiv-\sum_{{\sig}^{\A}}P_{\M}^{\A}({\sig}^{\A})\ln P_{\M}^{\A}({\sig}^{\A})$, is the key ingredient in the mutual information
estimator, which can also be written as $\mathcal{I}_{\A:\B}=S_{\M}^{\A}+S_{\M}^{\B}-S$. Using \eref{eq:PMA_three_term} the marginal entropy can be expressed as
\begin{equation}
\label{eq:SMA_gen}
 S_{\M}^{\A} = \ln Z+\big\langle \beta \sum_{i,j\in \A} \phi_{i,j}(\sig) \big\rangle_\HH +
\big\langle \ln Z^{\B}(\mathbf{q}({\sig}^{\A}),{\sig}^{\A}) \big\rangle_\HH + \mathcal{O}(1)
\end{equation}
where $\langle f(\sig) \rangle_\HH\equiv Z^{-1}\sum_{\sig}e^{-\beta\HH(\sig)}f(\sig)$ for a general function $f(\sig)$ and $\mathbf{q}(\sig)\equiv\mathbf{Q}(\sig)/l(\sig)$ with $l(\sig)$ denoting the number of spins in $\sig$. Using the above expression for $S$ and $S_{\M}^{\A}$ in \eref{eq:S_int} and \eref{eq:SMA_gen}, and the expression for $S_\M^\B$, obtained in a similar way as \eref{eq:SMA_gen}, the mutual information can be written as
\begin{eqnarray}
\label{eq:I_genA}
\fl\qquad I_{\A:\B}=\ln Z+\big\langle2\beta\sum_{i\in\A,j\in\B}\phi_{i,j}(\sig)\big\rangle_{\HH}-\big\langle\ln[Z^{\A}(\mathbf{q}({\sig}^{\B}),\sig^\B)Z^{\B}(\mathbf{q}({\sig}^{\A}),{\sig}^{\A})]\big\rangle_{\HH} \nonumber\\
\qquad\qquad\qquad\qquad-L\big\langle\beta\epsilon(\mathbf{q}(\sig))\big\rangle_{\HH}+\mathcal{O}(1).
\end{eqnarray}
The second term in the RHS of the above equation corresponds to the average of the short-range interaction term over the boundary of the bipartition. In one-dimension, the number of terms
in this sum does not increase with $L$, and it therefore contributes only to the $\mathcal{O}(1)$ term in $I_{\A:\B}$. The third and forth terms are evaluated below using the SPA.

In order to compute the term $\big\langle\ln[Z^{\A}(\mathbf{q}({\sig}^{\B}),\sig^\B)Z^{\B}(\mathbf{q}({\sig}^{\A}),{\sig}^{\A})]\big\rangle_{\HH}$,
 it is useful to consider the ensemble average of a general function of the $q$ variables in each of the
two subsystems, denoted by $\big\langle g({\bf q}({\sig}^{\A}),{\bf q}({\sig}^{\B})) \big\rangle_\HH$. Using the technique used above in the derivation of $Z$ \eref{eq:app_Z2}, the
average can be written as
\begin{eqnarray}
\fl  \quad  \big\langle g({\bf q}({\sig}^{\A}),{\bf q}({\sig}^{\B})) \big\rangle_\HH  = Z^{-1}\ell^{p}(L-\ell)^{p}\int d{\bf q}^{\A}d{\bf q}^{\B} \int d{\bf h}^{\A}d{\bf h}^{\B} \\
\fl  \qquad \quad \times g(\mathbf{q}^{\A},\mathbf{q}^{\B})e^{-L[\beta\epsilon(\alpha\mathbf{q}^{\A}+(1-\alpha)\mathbf{q}^{\B})+\alpha\mathbf{h}^{\A}\cdot\mathbf{q}^{\A}+(1-\alpha)\mathbf{h}^{\B}\cdot\mathbf{q}^{\B}]}\big\langle 1|T_{\beta,\mathbf{h}^{\A}}^{\ell}T_{\beta,\mathbf{h}^{\B}}^{L-\ell}|1\big\rangle\big[1+\mathcal{O}(L^{-1})\big].  \nonumber
\end{eqnarray}
 Evaluating using the SPA  the integrals over ${\bf h}^\A$ and ${\bf h}^\B$ of the leading order term in $L$ in $\big\langle g({\bf q}({\sig}^{\A}),{\bf q}({\sig}^{\B})) \big\rangle_\HH$ yields
\begin{equation}
\label{eq:avg_qa_qb}
\big\langle g({\bf q}({\sig}^{\A}),{\bf q}({\sig}^{\B})) \big\rangle_\HH  \simeq Z^{-1} \ell^{\frac{p}{2}}(L-\ell)^{\frac{p}{2}}\int d{\bf q}^{\A} d{\bf q}^{\B}g(\mathbf{q}^{\A},\mathbf{q}^{\B}) \omega(\mathbf{q}^{\A},\mathbf{q}^{\B}) e^{-L\beta f_{\HH}(\beta,\mathbf{q}^{\A},\mathbf{q}^{\B})},
\end{equation}
where
 \begin{equation}
f_{\HH}(\beta,{\bf q}^\A,{\bf q}^\B) =   \epsilon(\alpha \mathbf{q}^{\A}+(1-\alpha) \mathbf{q}^{\B})+\alpha f_\phi(\beta,\mathbf{q}^{\A})+(1-\alpha)f_\phi(\beta,\mathbf{q}^{\B}),
 \end{equation}
 is the Landau free energy of $({\bf q}^\A,{\bf q}^\B)$. The function
$\omega(\mathbf{q}^{\A},\mathbf{q}^{\B})$ accounts for the $\mathcal{O}(1)$ pre-factor of the leading order term in $\big\langle g({\bf q}({\sig}^{\A}),{\bf q}({\sig}^{\B})) \big\rangle_\HH$ and additional pre-factors that result from the SPA. For pure mean-field systems, where $\phi_{i,j}(\sig)=0$, it can be easily shown that $\omega(\mathbf{q}^{\A},\mathbf{q}^{\B})=\omega_0(\mathbf{q}^{\A})\omega_0(\mathbf{q}^{\B})$.

Using \eref{eq:avg_qa_qb} to evaluate the term $\big\langle\ln[Z^{\A}(\mathbf{q}({\sig}^{\B}),\sig^{\B})Z^{\B}(\mathbf{q}({\sig}^{\A}),\sig^{\A})]\big\rangle_{\HH}$ in \eref{eq:I_genA} yields different results depending whether the system is critical or not.  Away from criticality, the SPA of the integrals over ${\bf q}^\A$ and ${\bf q}^\B$ can be performed by expanding the
 exponential in \eref{eq:avg_qa_qb} to quadratic order in these variables. In this case since $\ln[Z^\A({\bf q}^\B,\sig^{\B}) Z^\B({\bf q}^\A,\sig^{\A})]$ is a slowly varying function in comparison to the exponential, its
leading order contribution involves only $\ln[Z^\A({\bf q}^\star,\sig^{\B}) Z^\B({\bf q}^\star,\sig^{\A})]$.
When inserting the result into \eref{eq:I_genA} the $\mathcal{O}(L)$ term cancels with that of $\big\langle\epsilon(\mathbf{q}(\sig))\big\rangle_{\HH}$, obtained in \eref{eq:app_S3}, yielding {\it away from criticality} the following result:
\begin{eqnarray}
 \label{eq:app_I_away}
\mathcal{I}_{\A:\B}= \frac{1}{2}\ln\left[ g(\alpha)g(1-\alpha)\right] + \mathcal{O}(1),
\end{eqnarray}
where $g(x)$ is in general a non-generic scaling function.

In order to derive $g$ it is useful to note that the only $\alpha$-dependent contribution to \eref{eq:app_I_away} comes from $(\det A_{1-\alpha})^{-1/2}$ term in \eref{eq:app_PMA}, which yields $g(x)=\det A_{1-x}$.
According to \eref{eq:app_f_def} and \eref{eq:app_A_def}, each element in $A(x)$ is a linear polynomial of $x$, whose coefficients depend in general
on the parameters of the model. The determinant, $\det A_x$, is thus a polynomial of the form $a_p x^p + a_{p-1} x^{p-1}+\ldots + a_1x +  a_0$.
 However, since the scaling function is determined up to a constant, it can be written in terms of the rescaled parameters $b_i=a_i/a_1$ as
\begin{equation}
\label{eq:gx_can}
g(x)= \det A_{1-x} =  b_p x^p + b_{p-1} x^{p-1}+\ldots + b_2 x^2 + x +b_0.
\end{equation}
This implies that $g(x)$ depends in fact only on $p$ parameters.

\begin{figure}
\begin{center}
\includegraphics[scale=0.6]{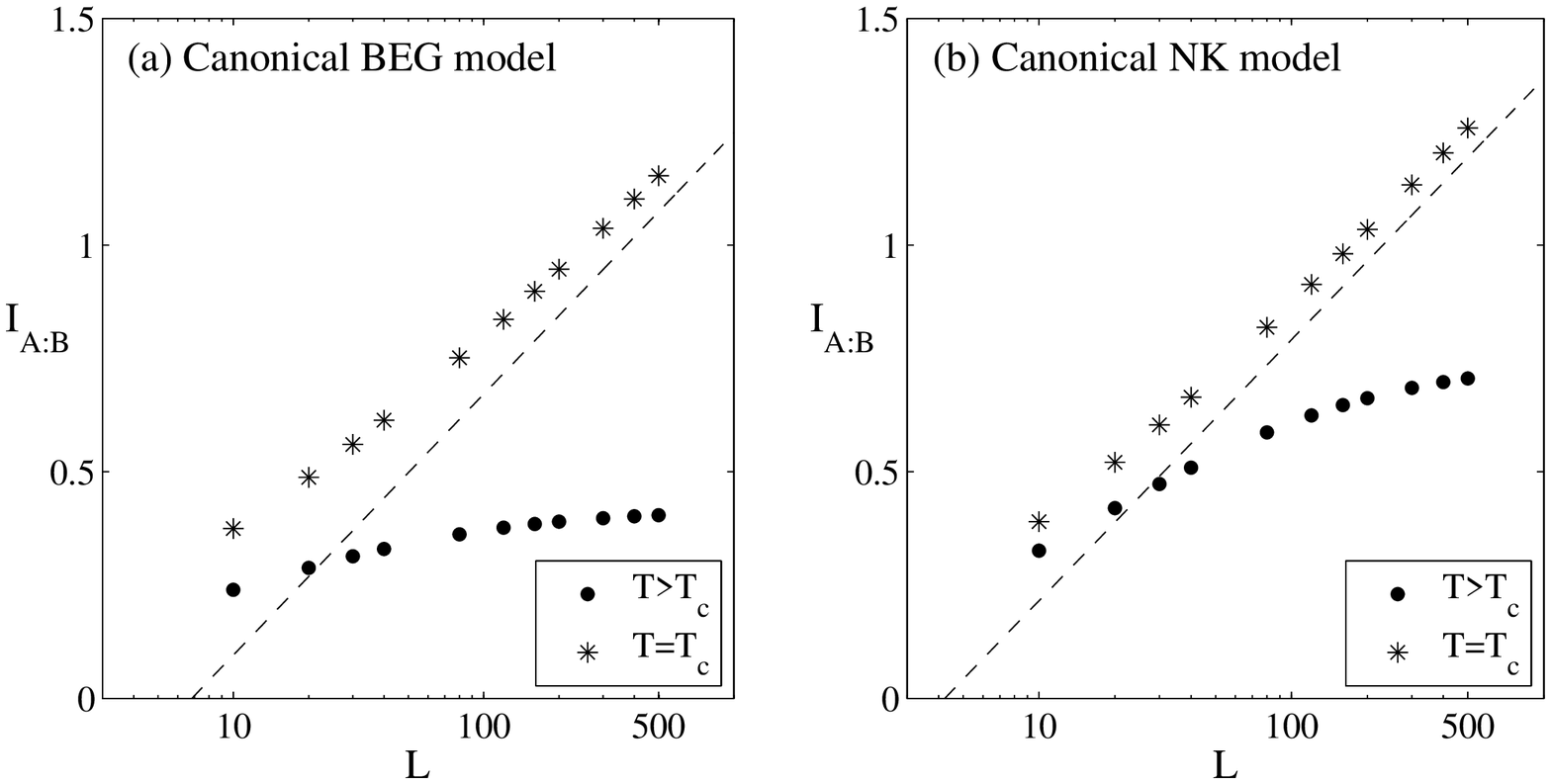}
\caption{
\label{fig:can_IAB}
Numerical evaluation of the mutual information estimator in the canonical ensemble for $\ell/L=1/2$, plotted as a function of the system-size, which is
given in a logarithmic scale. Figure (a) and (b) shows the results  for the BEG model and NK model, respectively.
The parameters used in the computation are identical to those described in the caption of \fref{fig:can_SAB}.
At criticality, the leading order term in $I_{\A:\B}$ is expected to scale as $\frac{1}{4}\log L$, denoted by the dashed lines, whereas away from criticality $I_{\A:\B}$ is expected to converge to a constant.}
\end{center}
\end{figure}

In general, one expects the leading term in $\mathcal{I}_{\A:\B}$ resulting from the SPA to scale as $\sqrt{L}$ at criticality. This is because $\mathcal{I}_{\A:\B}$
 involves the term $L\big\langle\epsilon(\mathbf{q}(\sig))\big\rangle_{\HH}$ which leads to a divergence of this kind in the case of
$\mathcal{S}_{\A:\B}$. In \ref{app:vanish} it shown, however, that the $\sqrt{L}$ term that comes from the energy
cancels exactly with the one that come from $\big\langle\ln[Z^{\A}(\mathbf{q}({\sig}^{\B}))Z^{\B}(\mathbf{q}({\sig}^{\A}))]\big\rangle_{\HH}$.
The remaining leading order term in \eref{eq:I_genA} comes from the $\ln Z$. Inserting the form of $Z$ in (\ref{eq:app_Z_at}) yields {\it at criticality}
 \begin{eqnarray}
 \label{eq:app_I_at}
\mathcal{I}_{\A:\B}=\frac{1}{4}\ln L+ \frac{1}{2}\ln\left[ g(\alpha)g(1-\alpha)\right]+
\mathcal{O}\left( 1 \right).
 \end{eqnarray}
The results in \eref{eq:app_I_away} and \eref{eq:app_I_at} are verified numerically in \fref{fig:can_IAB} for the BEG and the NK models.

At criticality, since $f_{\HH,1}$ is the landau free energy of the whole system, we find that $\det A_1=0$. This implies in turn that $b_0=0$ in \eref{eq:gx_can},
and that $g(x)$ involves only on $p-1$ non-generic parameters. As a result, for $p=1$, such as in the NK model, $g$ has a generic form, $g(x)=x$.
It is interesting to note that $b_p$ is proportional to the determinant of the Hessian (discriminant) of $\epsilon({\bf q})$, $b_p = a_1^{-1} \det (  \partial ^2 \epsilon / \partial q_i\partial q_j \vert_{{\bf q}^\star} )$.
In the BEG model, where $p=2$, one would expect to obtain the a non-generic scaling function of the form $g(x)=b_2 x^2+ x $.
However, the fact that the discriminant of the energy vanishes at criticality leads to a generic scaling function, $g(x)=x$.

\subsection{Micro-canonical ensemble}
\label{sec:gen_micro}
In this section we study the behavior of the information estimators in the generic model, defined in \eref{eq:Ham_generic}, within the micro-canonical ensemble, where the total energy is fixed.
The main results of this calculation are summarized in \sref{sec:sum_mc}.
The computation is done by following the lines of derivation presented in the previous section, while omitting several of the steps for the sake of brevity.

\subsubsection{Degeneracy:}
 In the micro-canonical ensemble the probability distribution is uniform over all spin configurations with a certain energy, $E$. Mathematically this can be written as
\begin{equation}
P\left(\sig\right)=\Omega^{-1} \int_{E-\delta /2}^{E+\delta/2}dE'\delta(E'- \HH( \sig,L)),
\end{equation}
where $\delta\sim \mathcal{O}(1)$ is a finite parameter and $\Omega=L\int_{E-\delta/2}^{E+\delta/2} dE'\sum_{\sig}\delta(E'-\HH(\sig))$ is the number of micro-states with
energy between $E-\delta/2$ and $E+\delta/2$.

Similarly to the derivation of $Z$ in the previous section, it is useful to express the degeneracy, $\Omega$, in terms of an integral over continuous variable as
\begin{equation}
\label{eq:OmegaE_00}
\Omega=L^{p}\int dE'\int d{\bf q} \sum_{\sig} \delta\big[E'-\sum_{i,j}\phi_{i,j}(\sig)-L\epsilon(\mathbf{q})\big]\prod_{j=1}^{p}\delta( Q_{j}(\sig)-L q_j).
\end{equation}
As before, the delta function can be replaced by an integral over the fields ${\bf h}$, yielding
\begin{equation}
\label{eq:OmegaE_0}
\Omega=L^{p}\int d{\bf q} d{\bf h} dE' d\beta e^{\beta E'-L\beta\epsilon(\mathbf{q})}\langle 1 | T_{\beta,\mathbf{h}}^{L} |1\rangle,
\end{equation}
where $\langle 1 | T_{\beta,\mathbf{h}}^{L} |1\rangle$ denotes as in \eref{eq:Z_can_1} the partition function of a short-range interacting systems whose Hamilton
is given in \eref{eq:Ham_short}.

Because the integrand in \eref{eq:OmegaE_0} does not vary significantly in the interval $E'\in [E-\delta/2,E+\delta/2]$, the integral over $E'$ can be replace by the value of the integrand at $E'=E$. This would result in errors that
scale as $e^{-\beta\delta}$ which can be written as $\mathcal{O}(1)$.
The integral over  $\beta$, on the other hand, has to be evaluated using the SPA, yielding
\begin{equation}
\label{eq:OmegaE_A}
\Omega=L^{p/2-1/2}\int d{\bf q} e^{-L y_\HH(E/L,{\bf q})}\omega(\mathbf{q})[\mathcal{O}(1)+\mathcal{O}(L^{-1})],
\end{equation}
where $y_\HH(\varepsilon,{\bf q})\equiv \beta^\star(\varepsilon,{\bf q}) [ \varepsilon - f_\HH (\beta^\star(\varepsilon,{\bf q}),\mathbf{q})]$ can be regarded as the Landau free energy of the micro-canonical system and $f_\HH$ is defined below \eref{eq:app_Z2}. Here $\beta^\star(\varepsilon,{\bf q})$ is the saddle point of the integral over $\beta$, defined via the equation
\begin{equation}
\phi(\beta^\star,{\bf q})=\varepsilon-\epsilon (\mathbf{q}).
\end{equation}
At this inverse temperature the average energy in the short-range interacting system is equal to the difference between the overall energy and the mean-field energy.

In the case of a pure mean-field system, where $\phi_{i,j}(\sig)=0$, the integral over $\beta$ can be replaced by a delta function, yielding
\begin{equation}
\label{eq:OmegaE_B}
\Omega=L^{p/2-1}\int d{\bf q} \delta(\epsilon (\mathbf{q})-E/L) \omega(\mathbf{q})[\mathcal{O}(1)+\mathcal{O}(L^{-1})].
\end{equation}
As expected, in the absence of additional short-range interactions, the mean-field energy is strictly fixed, $\epsilon (\mathbf{q})=E/L$.
The derivations of the information estimators for $\phi_{i,j}(\sig)\neq0$ and for $\phi_{i,j}(\sig)=0$ are slightly
different, as indicated by the difference between \eref{eq:OmegaE_A} and \eref{eq:OmegaE_B}. In both cases, however, one finds the same leading order scaling of $\mathcal{S}_{\A:\B}$ and  $\mathcal{I}_{\A:\B}$. For brevity, we present only the analysis of the more general case where $\phi_{i,j}(\sig)\neq0$.

As in the case of the analysis of $Z$ in the previous section, the integral in \eref{eq:OmegaE_A} can be evaluated using the SPA, which is performed in
two steps. In the first step we approximate the integrals over
 $q_2,\dots,q_p$, by expanding the exponent to quadratic order in these variables. The next step is to approximate the remaining one-dimensional
 integral, by expanding the exponent to order $q_1^2$ away from criticality and to order $q_1^4$ at criticality.
  The first step of the SPA yields,
 \begin{equation}
 \label{eq:Omega_first_step}
   \Omega =  \int dq_1 \tilde{\omega}(q_1) e^{ -L y_\HH(E/L,\tilde{{\bf q}})}\big[\mathcal{O}(1)+\mathcal{O}(L^{-1})\big]
  \end{equation}
 where in this section $\tilde{\bf q}=\left(q_1,{\tilde q}_2(q_1),{\tilde q}_3(q_1),\dots, {\tilde q}_p(q_1)\right)$ is the solution of the set of equations
 $\partial y_\HH/ \partial q_i =0$ for $i=2,3,\dots,p$.
The function $\tilde{\omega}(q_1)$ accounts for the contribution from $\omega (\tilde{{\bf q}})$ and the coefficients that results from the SPA and thus
  \begin{equation}
\label{eq:app_omega_tilde}
   \tilde{\omega}(q_1)\equiv \omega (\tilde{{\bf q}}(q_1)) \pi^{(p-1)/2} \Big[ \det \Big( \frac{\beta \partial^2 y_\HH(E/L,{\bf q})}{\partial q_i \partial q_j} \Big\vert_{\tilde{{\bf q}}(q_1)} \Big)\Big]^{-1/2},
  \end{equation}
 for $i,j=2,3,\dots,p$.

{\it Away from criticality} the second step of the SPA yields
\begin{eqnarray}
\label{eq:gen_Omega_away}
 \Omega \simeq e^{L s_\phi(\frac{E}{L},{\bf q}^\star)} L^{-1/2}\tilde{\omega}(q_1^\star)  \chi_{0,2} \Big( \frac{d^2 y_\HH (\frac{E}{L},\tilde{{\bf q}}(q_1))}{d q_1^2} \Big\vert_{q^\star_{1}} \Big),
\end{eqnarray}
where $s_\phi(\varepsilon,{\bf q}) = \beta\big[ f_\phi (\beta^\star(\varepsilon,{\bf q}),\mathbf{q}) - \phi(\beta^\star(\varepsilon,{\bf q}),\mathbf{q})\big]$ is the entropy
of the short-range interacting system, described by $\HH'$ in \eref{eq:Ham_short}, for $\beta=\beta^\star(\varepsilon,{\bf q})$ and with ${\bf h}$ set such that $\langle{\bf Q}(\sig)\rangle_{\HH'}/L={\bf q}$. The point ${\bf q}={\bf q}^\star$ is the global minimum of $y_\HH(\frac{E}{L},{\bf q})$. Since for ${\bf q}={\bf q}^\star$, the energy terms in $y_\HH$ cancel, yielding $y_\HH(\varepsilon,{\bf q}^\star)= -s_\phi(\varepsilon,{\bf q}^\star)$, one can use $s_\phi$ in the exponent
in \eref{eq:gen_Omega_away}.
 As in the derivation of $Z$ in \eref{eq:app_Z_at}, {\it at criticality} the SPA of the integral in \eref{eq:OmegaE_A}
 yields a different polynomial-dependence in $L$, given by
\begin{eqnarray}
\label{eq:gen_Omega_at}
 \Omega \simeq e^{L s_\phi(\frac{E}{L},{\bf q}^\star)} L^{-1/4}\tilde{\omega}(q_1^\star)  \chi_{0,4} \Big( \frac{d^4 y_\HH (\frac{E}{L},\tilde{{\bf q}}(q_1))}{d q_1^4} \Big\vert_{q^\star_{1}} \Big).
\end{eqnarray}

\subsubsection{Separation entropy:} In order to compute the separation entropy, one has to compute first the Shannon entropy of the whole system, which in
the micro-canonical ensemble is given simply by $S= \ln \Omega$.
Here we consider the separation process discussed in the case of the canonical ensemble, whereby the Hamiltonians of the separated subsystems are given by \eref{eq:Ham_generic} with
$L$ replaced by the corresponding length of each subsystem. This assures that the
average values of ${\bf q}$ of the separated subsystems are identical to those of the composite system.
Since the decoupled subsystems maintain the form of the Hamiltonian of the whole system, their entropies are given by $S=\ln \Omega$ where $\Omega$ is given by \eref{eq:gen_Omega_away} and
 \eref{eq:gen_Omega_at} with
$L$ replaced by $\ell$ for subsystem $\A$ and by $L-\ell$ for subsystem $\B$.

As a result the extensive terms in the separation entropy cancel and the remaining leading order terms are given {\it away from criticality} by
\begin{eqnarray}
\label{eq:BEG_micro_Omega_away}
S_{\A:\B}=\frac{1}{2}\ln\left[ L \alpha \left( 1-\alpha \right)
\right]+\mathcal{O}\left( 1 \right),
\end{eqnarray}
and {\it at criticality} they are equal to
\begin{eqnarray}
\label{eq:BEG_micro_Omega_at}
S_{\A:\B}=\frac{1}{4}\ln\left[ L \alpha \left( 1-\alpha \right)
\right]+\mathcal{O}\left( 1 \right).
\end{eqnarray}
This scaling form of $S_{\A:\B}$ is identical to that obtained in the canonical ensemble, up to an addition of a $\frac{1}{2}\ln L$ term both at criticality
and away from criticality. This term is due to the fixed energy constraint, $L \epsilon({\bf Q}(\sig)/L)+\sum_{i,j} \phi_{i,j}(\sig) = E$, which introduces additional correlations between the spin variables.

\begin{figure}
\begin{center}
\includegraphics[scale=0.6]{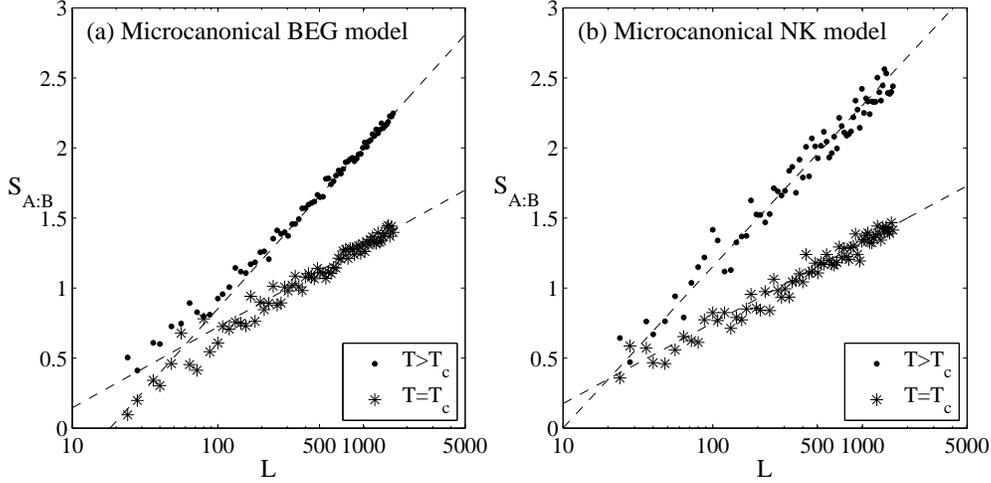}
\caption{
\label{fig:mc_SAB} Numerical evaluation of the Shannon separation estimator in the micro-canonical ensemble for $\ell/L=1/2$, plotted as a function of the system-size, given in a logarithmic scale.
Figure (a) shows the results for the BEG model at criticality ($\Delta/J\approx 0.347$, $E/L\Delta= 0.5$) and away from criticality ($\Delta/J\approx 0.530$, $E/L\Delta= 0.5$), denoted by $\star$ and $\bullet$ respectively.
Similarly, figure (b) shows the results for the NK model at criticality ($K/J\approx -0.333$, $E/L K = 0.65$) and away from criticality ($K/J\approx -0.5$, $E/L K = 0.65$).
At criticality and away from criticality, the leading order term in $S_{\A:\B}$ is expected to scale as $\frac{1}{4}\log{L}$ and  $\frac{1}{2}\log{L}$, respectively. These $\log L$ terms are denoted by the straight dashed lines.}
\end{center}
\end{figure}

Similarly to the canonical case, these results can be verified numerically, as shown in \fref{fig:mc_SAB}.
 Here, however, the constant term in $S_{\A:\B}$ was found to oscillate with some finite scale. A convincing fit thus required sampling a large number of system sizes.
In order to avoid the arbitrariness in value of the parameter $\delta$, the integral over $E'$ in \eref{eq:OmegaE_00} was performed numerically over $E'\in (-\infty,E]$ instead of
$E'\in[E-\delta/2,E+\delta/2]$. These two definitions of the micro-canonical ensemble can be shown in our case to yield the same scaling form of $S_{\A:\B}$ as well as of $I_{\A:\B}$, computed
below.

\subsubsection{Mutual information}
The mutual information is computed from the marginal probability distribution. In the micro-canonical ensemble the latter is given by
a sum over microstates with a energy between $E-\delta/2$ and $E+\delta/2$, which can be written as
\begin{eqnarray}
\label{eq:gen_pm_micro}
P_{\M}^{\A}({\sig}^{\A})&=&\Omega^{-1}\int_{E-\delta/2}^{E+\delta/2} dE'\sum_{\sig^{\B}}\delta(E'-\HH({\sig}^{\A},{\sig}^{\B}))  \\ &=&\Omega^{-1}Z^{\B}\Big(\varepsilon,\mathbf{q}({\sig}^{\A}),\sig^{\A}\Big)\mathcal{O}(1), \nonumber
\end{eqnarray}
where $\varepsilon=\frac{E}{L}-\frac{1}{L}\sum_{i,j\in\A}\phi_{i,j}(\sig^\A)$ and the partition function over subsystem $\B$ is defined in this section as
\begin{equation}
Z^{\B}(\varepsilon,\mathbf{q}^{\A},\sig^{\A})\equiv \int d\beta e^{L\beta \varepsilon }\sum_{{\sig}^{\B}}e^{-L\beta\epsilon(\mathbf{q}({\sig}^{\A}),\mathbf{q}({\sig}^{\B}))-\beta\sum_{i,j\in\B}\phi_{i,j}(\sig)}e^{-2\beta\sum_{i\in\A,j\in\B}\phi_{i,j}(\sig)}.
\end{equation}
The $\mathcal{O}(1)$ term in \eref{eq:gen_pm_micro} comes from the approximation of the integral over $E'$ by a delta function at $E'=E$.

As in the derivation of $Z^\B$ in the canonical ensemble in \eref{eq:ZBa}, one can introduce an integral over the $q$ variables of subsystem
$\B$ and replace the resulting delta function by an integral over ${\bf h}$, yielding
\begin{equation}
Z^{\B}(\varepsilon,\mathbf{q}^{\A},\sig^{\A})=L^{p}\int d{\bf q}^{\B}d{\bf h}^{\B}d\beta e^{\beta L\epsilon-L\beta\epsilon(\mathbf{q}^{\A},
\mathbf{q}^{\B})}\big\langle \sig^{\A}_{\mathrm{b}}|T_{\beta,\mathbf{h}^{\B}}^{L-\ell}|1\big\rangle.
\end{equation}
Considering only the leading order contribution in $L$ to $\big\langle \sig^{\A}_{\mathrm{b}}|T_{\beta,\mathbf{h}^{\B}}^{L-\ell}|1\big\rangle$ and evaluating the integral
over ${\bf h}^\B$ and $\beta$ using the SPA yields
\begin{equation}
\label{eq:gen_micro_zB}
Z^{\B}(\varepsilon,\mathbf{q}^{\A},\sig^{\A})=L^{(p-1)/2}\int d{\bf q}^{\B}e^{-L  y_{\HH,1-\alpha}(\varepsilon,\mathbf{q}^{\A},\mathbf{q}^{\B})  }\omega(\mathbf{q}^{\B},\sig^{\A})
\big[1+\mathcal{O}(L^{-1})\big],
\end{equation}
where
\begin{equation}
 y_{\HH,x}(\varepsilon,\mathbf{q}^{\A},\mathbf{q}^{\B}) =  \beta^{\star}_x[\varepsilon-x f_{\HH,x}(\beta^{\star}_x,\mathbf{q}^{\A},\mathbf{q}^{\B})].
\end{equation}
Here $\beta^\star_x$  is the saddle point of the integral over $\beta$, defined via the equation $x \phi(\beta^\star_x,{\bf q}^\B)=\varepsilon-\epsilon (\mathbf{q}^\A,\mathbf{q}^\B)$.
The function $y_{\HH,x}(\epsilon,{\bf q}^\A,{\bf q}^\B)$ can be regarded as the Landau free energy of a subsystem of size $xL$, given the values of the $q$ variables in the complementary subsystem, denoted by ${\bf q}^\A$.

Using \eref{eq:gen_pm_micro} the mutual information estimator can be written as
\begin{equation}
\label{eq:micro_IAB}
\mathcal{I}_{\A:\B}=  \ln \Omega -\big\langle \ln \big[Z^{\B}(\frac{E}{L}-\phi^\A,\mathbf{q}^{\A},\sig^{\A}) Z^{\A}(\frac{E}{L}-\phi^\B,\mathbf{q}^{\B},\sig^{\B})\big]\big\rangle_{\HH} + \mathcal{O}(1),
\end{equation}
where $\phi^k\equiv\frac{1}{L}\sum_{i,j\in k} \phi_{i,j}(\sig)$ for $k=\A,\B$. Following the same reasoning described in \ref{app:vanish} it can be shown that the second term in RHS above does not yield a $\sqrt{L}$-divergence at criticality and that its leading order contribution the same as that of
$\big\langle \ln \big[Z^{\B}(\frac{E}{L}-\alpha\phi(\beta^\star,{\bf q}^\star),{\bf q}^\star,\sig^{\A}) Z^{\A}(\frac{E}{L}-(1-\alpha)\phi(\beta^\star,{\bf q}^\star),{\bf q}^\star,\sig^{\B})\big]\big\rangle_{\HH}$. For subsystem $\B$ one can compute $Z^{\B}(\frac{E}{L}-\alpha\phi(\beta^\star,{\bf q}^\star),{\bf q}^\star,\sig^{\A})$
by evaluating the integral in \eref{eq:gen_micro_zB} using the SPA which yields
\begin{equation}
\label{eq:gen_micro_zBstar}
Z^{\B}(\frac{E}{L}-\alpha \phi(\beta^\star,{\bf q}^\star),{\bf q}^\star,\sig^{\A}) =
L^{-1/2}\omega({\bf q}^\star,\sig^\A) \pi^{p/2} e^{L\alpha s_\phi(\frac{E}{L},{\bf q}^\star)}\big[1+\mathcal{O}(L^{-1})\big],
\end{equation}
where in this section the Hessian matrix is defined as
\begin{equation}
\label{eq:app_A_def1}
(A_x)_{i,j}= \beta \frac{\partial ^2  y_{\HH,x} (E/L-\phi(\beta^\star,\mathbf{q}^{\star}),\mathbf{q}^{\star}, \mathbf{q}^\B) }{\partial q_i^\B \partial q_j^\B} \Big\vert_{{\bf q}^\B={\bf q}^\star}.
\end{equation}
The same expression for $Z^\A$ is obtained by replacing $\alpha \to 1 - \alpha$ and exchanging $\A$ and $\B$ in \eref{eq:gen_micro_zBstar}.

\begin{figure}
\begin{center}
\includegraphics[scale=0.6]{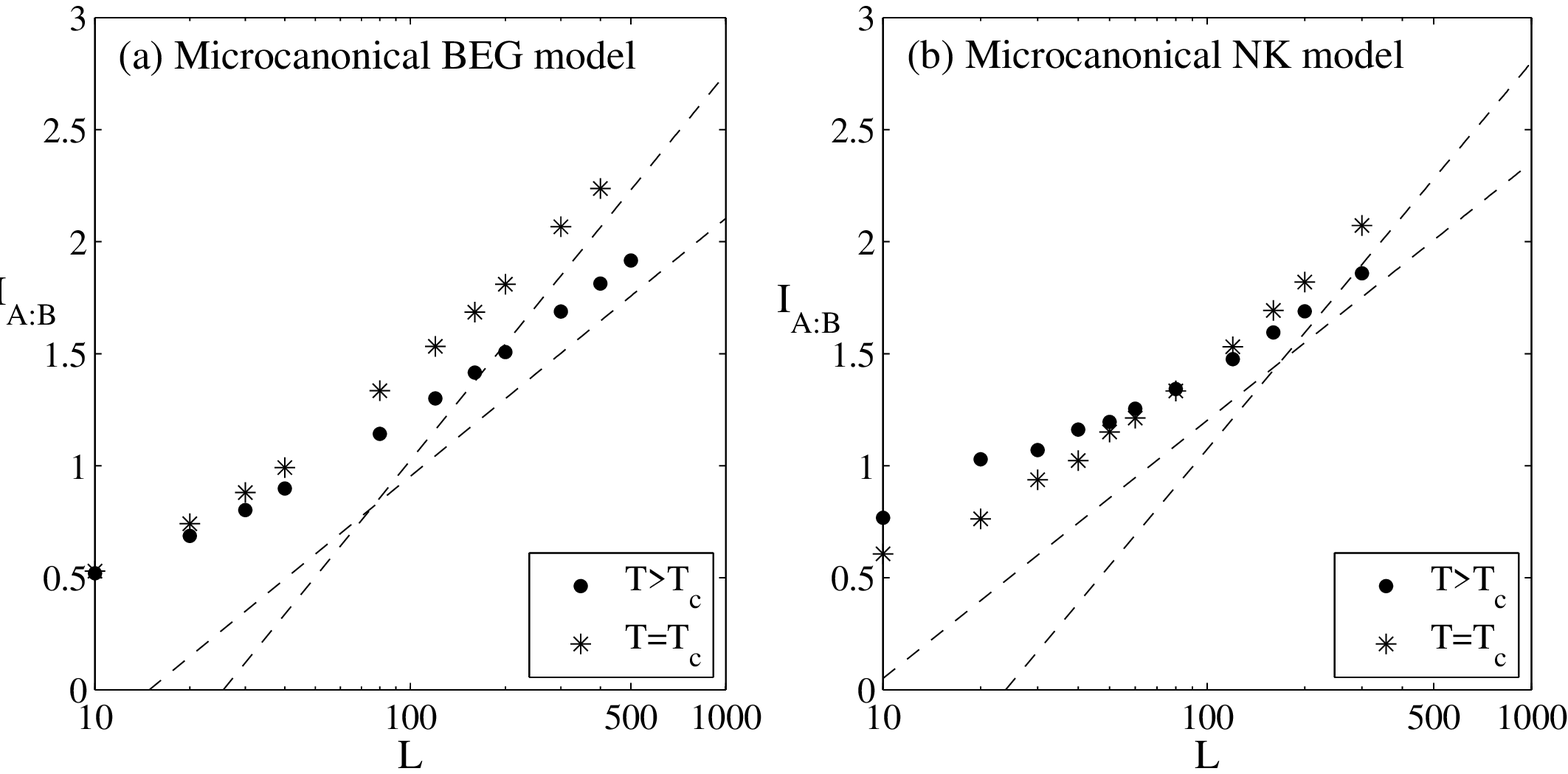}
\caption{
\label{fig:mc_IAB} Numerical evaluation of the mutual information estimator in the canonical ensemble for $\ell/L=1/2$, plotted as a function of the system-size. Figure (a) and (b) shows the results  for the BEG model and NK model, respectively.
The parameters used in the computation are identical to those described in the caption of \fref{fig:mc_SAB}.
At criticality and away from criticality, the leading order term in $I_{\A:\B}$ is expected to scale as $\frac{3}{4}\log{L}$ and  $\frac{1}{2}\log{L}$, respectively.  These $\log L$ terms are denoted by the straight dashed lines.}
\end{center}
\end{figure}

Inserting \eref{eq:gen_micro_zBstar} and the expression for $\Omega$ in (\ref{eq:BEG_micro_Omega_away}) and (\ref{eq:BEG_micro_Omega_at}) into \eref{eq:micro_IAB}
yields {\it away from criticality} the following scaling form:
\begin{eqnarray}
 \label{eq:micro_I_away}
\mathcal{I}_{\A:\B}= \frac{1}{2}\ln L  + \frac{1}{2} \ln [g(\alpha)g(1-\alpha)] + \mathcal{O}(1),
 \end{eqnarray}
 whereas {\it at criticality} the mutual information is given by
\begin{eqnarray}
 \label{eq:micro_I_at}
\mathcal{I}_{\A:\B}=\frac{3}{4}\ln L  + \frac{1}{2} \ln [g(\alpha)g(1-\alpha)]+\mathcal{O}\left( 1 \right).
 \end{eqnarray}
 Here $g(x)=\det A_{x}$ with $A_x$ defined in \eref{eq:app_A_def1}. Similarly to the canonical case, these results can be verified numerically, as shown in \fref{fig:mc_IAB}.
 Here, however, the mutual information appears to converge more slowly with $L$ than in the canonical ensemble.

 As in the canonical ensemble, one can show that $g(x)$ is a polynomial of degree $p$, of the form given in \eref{eq:gx_can},
 and that at criticality $b_0=0$. This implies that $g(x)$ depends on $p-1$ parameters at criticality and on $p$ parameters away from criticality.
For $p=1$ such as in the NK model, $g$ therefore has a generic form at criticality, $g(x)=x$.
In the case of pure mean-field models, the fixed energy constraint reduces the dimension of $A(x)$ to $(p-1)\times(p-1)$, which implies that in this case
$g(x)$ is a polynomial of degree $p-1$. At criticality, $g(x)$ therefore depends on $p-2$ parameters in pure mean-field model. This implies that the BEG model, which is a pure mean-field model with $p=2$, also exhibits at criticality a generic scaling function of the form $g(x)=x$. The scaling function, $g(\alpha)g(1-\alpha)$, of the BEG model is plotted for several temperatures in \fref{fig:scaling_I}.

\subsection{Ground-state degeneracy and first order transitions}
\label{sec:beg_degeneracy}
In all the above analysis the Landau free energies, $f_\HH(\beta,{\bf q})$ in the canonical ensemble and $y_\HH(E/L,{\bf q})$ in the microcanonical ensemble,
were assumed to have a single ground state, denoted by ${\bf q}^\star$.
However, specific symmetry properties may lead to degenerate ground states. This is true in the BEG model and the NK model, where the spin flip symmetry, $\sig \to -\sig$,
yields two degenerate ordered states with opposite magnetization. Even models without symmetry exhibit degenerate ground state at first order phase transition points.
 From a dynamical point, in a mean-field system the tunneling time between these ground-states grows exponentially with $L$.
The effect of having multiple ground states on the information estimators is therefore
relevant for finite systems that are measured over a very long period of time. Nevertheless, this degeneracy can be taken into account in the above derivation, as demonstrated in this section. The results show that the degeneracy affects only the $\mathcal{O}(1)$ terms of the
 information estimators. Specifically, this implies that $\mathcal{S}_{\A:\B}$ and $\mathcal{I}_{\A:\B}$ remain finite at first order transitions points.

 We demonstrate the effect of the ground-state degeneracy within the canonical ensemble,
 where we denote the global minima of $f_\HH(\beta,{\bf q})$ by ${\bf q}^{\star \,(j)}$ with $j=1,\ldots,K$ and $K$ being the number of degenerate states.
The first calculation in \sref{sec:gen_can} to be affected by this degeneracy is the evaluation the integral in $Z$,
\begin{equation}
 Z=L^{p/2}\int d{\bf q} \omega(\mathbf{q}) e^{-L\beta f_\HH(\beta,{\bf q})}\big[1+\mathcal{O}(L^{-1})\big],
\end{equation}
given in \eref{eq:app_Z2} and rewritten here for convenience. In the case of degeneracy the SPA of the integral over ${\bf q}$ yields,
\begin{equation}
 Z  = \sum_{j=1}^K \tilde{\omega}(q^{\star\,(j)}_{1}) e^{-L \beta f_\HH(\beta,{\bf q}^{\star\,(j)})}  \chi_{0,2}\left( \beta \frac{d^2 f_\HH(\tilde{{\bf q}}(q_1))}{d q_1^2} \Big\vert_{q^{\star\,(j)}_{1}} \right)\big[1+\mathcal{O}(L^{-1})\big],
\end{equation}
which is in fact a sum over the degenerate ground states of the result obtained in the case of a single ground state (\ref{eq:app_Z_away}).
It is easy to show that a similar sum appears in the expression for $S$ and thus also in the expression
for $S_{\A:\B}$. This eventually yields $S_{\A:\B}=\mathcal{O}(1)$ away from criticality and on the first order transition line.

At criticality the sum of $S_{\A:\B}$ over the different minima would yield
the same form as in the case of a single ground state, only with a different coefficient in front of the $\sqrt{L}$ term, given by
\begin{eqnarray}
\label{eq:gamma details}
\gamma &=& \sum_{j=1}^K \beta \frac{\chi_{2,4}(\beta\frac{d^4 f_{\HH}}{d q^4_1}|_{q_1^{\star\,(j)}})}{\chi_{0,4} (\beta\frac{d^4 f_{\HH}}{d q^4_1}|_{q_1^{\star\,(j)}}) \omega({\bf q}^{\star,(j)})} \\
&& \times \frac{d}{dq_1^2} \big\{ \tilde{\omega}(q_1)[\epsilon( \tilde{\bf q}(q_1))+\phi(\beta, \tilde{\bf q}(q_1)) - \epsilon({\bf q}^{\star,(j)})-\phi(\beta,{\bf q}^{\star,(j)}) ] \big\} \Big|_{q_1^{\star\,(j)}}.\nonumber
 \end{eqnarray}
It is important to note, however, that in simple models one does not expect to find degenerate ground state at criticality. This would happen in models that have several symmetries that are broken at different critical points.

Using the same reasoning in the calculation of the mutual information, the term
$\big\langle\ln[Z^{\A}(\mathbf{q}({\sig}^{\B}),\sig^\B)Z^{\B}(\mathbf{q}({\sig}^{\A}),{\sig}^{\A})]\big\rangle_{\HH}-L\big\langle\epsilon(\mathbf{q}(\sig))\big\rangle_{\HH}$
in \eref{eq:I_genA} can be written as a sum over the degenerate ground states. This sum involves the value of $\ln Z^{\B}(\mathbf{q}^{\star,(j)},{\sig}^{\A})$ and
$\ln Z^{\A}(\mathbf{q}^{\star(j)},{\sig}^{\B})$. The evaluation of latter expressions using the SPA does not yield a sum over the degenerate ground-states. This can be understood by writing
 $Z^{\B}(\mathbf{q}^{\star\,(j)},\sig^\A)$ explicitly as
\begin{equation}
\label{eq:ZB_B}
Z^{\B}(\mathbf{q}^{\star\,(j)},\sig^\A)=(L-\ell)^{p/2}\int d{\bf q}^{\B}e^{-(L-\ell)\beta f_{\HH,1-\alpha}(\beta,\mathbf{q}^{\star\,(j)},{\bf q}^\B)}\omega(\mathbf{q}^{\B},\sig^\A)
\big[1+\mathcal{O}(L^{-1})\big],
\end{equation}
where $f_{\HH,1-\alpha}(\beta,\mathbf{q}^{\star\,(j)},{\bf q}^\B) \equiv \frac{1}{1-\alpha}\epsilon( \alpha \mathbf{q}^{\star\,(j)}+ (1-\alpha) \mathbf{q}^{\B})+f_\phi(\beta,\mathbf{q}^{\B})$. Using the fact that $\mathbf{q}^{\star\,(j)}$ are minima of $f_\HH(\mathbf{q})=\epsilon(\mathbf{q})+f_\phi(\mathbf{q})$
it is easy to see that only $\mathbf{q}^{\star\,(j)}$ minimizes $f_{\HH,1-\alpha}(\beta,\mathbf{q}^{\star\,(j)},{\bf q}^\B)$.
 Physically this implies that the value of ${\bf q}$ in subsystem $\A$ has broken the symmetry of
the free energy in $\B$. As a result the SPA of \eref{eq:ZB_B} involves only $\mathbf{q}^{\star\,(j)}$. Using a similar argument for $Z^{\A}(\mathbf{q}^{\star,(j)},{\sig}^{\B})$ one can show that
 that $I_{\A:\B}$ is also given by the same scaling forms obtained in the previous sections.

\subsection{Small $\ell/L$ scaling}
\label{sec:beg_finite}
The scaling forms of the information estimators, obtained in \sref{sec:gen_can} and \sref{sec:gen_micro}, have been derived for the case where each subsystem, $\A$ and $\B$, comprises
 a finite fraction of the entire system, i.e. $\ell\sim L$.
It is interesting to study how the scaling of the information estimators changes in the case where one of the subsystems, chosen here to be $\A$, is much smaller than the other but is still very large (allowing
SPA to be employed). We denote this limit as $1\ll \ell \ll L$.

As discussed above, the separation entropy is given by $S_{\A:\B}=S(L)-S(\ell)-S(L-\ell)$, where $S(x)$ denotes the entropy of a system of size $x$.
The expression of $S(x)$, computed in sections \ref{sec:gen_can} and \ref{sec:gen_micro} for the generic model, remains valid for any $x\gg1$. Inserting the expressions of $S(x)$ into
the $S_{\A:\B}$ is equivalent to taking the $\alpha \to 0$ limit in the expressions for $S_{\A:\B}$ obtained above.
The results of this calculation are summarized in \sref{sec:sum_finite_size} and not repeated here for the sake of brevity.

 This approach cannot be applied in the case of the mutual information, where the $L\to\infty$ and $\alpha\to0$ limit do not commute.
 In the canonical ensemble $\mathcal{I}_{\A:\B}$ was found to be given by
 $\mathcal{I}_{\A:\B}=\frac{1}{4}\ln L+ \frac{1}{2}\ln[ g(\alpha)g(1-\alpha)]+\mathcal{O}( 1 )$, where $g(x)$ is a non-generic polynomial.
 At criticality, the constant term in $g(x)$ vanishes and hence $g(\alpha)= b_1 \alpha+O(\alpha^2)$ for $\alpha\ll1$.
 This yields $\mathcal{I}_{\A:\B} \sim \frac{1}{4}\ln\frac{\ell^2}{L}$, which diverges to $-\infty$ with $L$ for $\ell \ll L^{1/2}$. This is in contrast with the intuition that the mutual information should be positive and diverge only with $\ell$ for $\ell \ll L$.
The term $\ln\frac{\ell^2}{L}$ suggests that there exists an intermediate scale, $\ell\sim L^{1/2}$, where the derivation in sections \ref{sec:gen_can} and \ref{sec:gen_micro} ceases to be correct.
In order to show this explicitly and obtain the correct scaling of $I_{\A:\B}$ we consider limit where
\begin{equation}
\ell= a  L^\zeta, \qquad 0<\zeta<1.
\end{equation}

We begin the analysis of $I_{\A:\B}$ from \eref{eq:I_genA}, which, by using the result of \ref{app:vanish}, can be written as
\begin{equation}
\label{eq:I_genA1}
I_{\A:\B}=\ln Z-\big\langle\ln[Z^{\A}(\mathbf{q}^\star,\sig^\B)Z^{\B}(\mathbf{q}^\star,{\sig}^{\A})]\big\rangle_{\HH}-L\big\langle\beta\epsilon(\mathbf{q}(\sig))\big\rangle_{\HH}+\mathcal{O}(1).
\end{equation}
We analyze first the expression for $Z^{\A}$. Starting from \eref{eq:app_Z2}, it can be evaluated the SPA with the exponentials expanded to quadratic order, yielding
\begin{eqnarray}
 Z^{\A}(\mathbf{q}^{\star},\sig^\B) &=& \ell^{p/2}\int d{\bf q}^{\A}e^{-\ell\beta f_{\HH,\alpha}(\beta,{\bf q}^{\star},{\bf q}^\A)}\omega(\mathbf{q}^{\A},\sig^\B)\big[1+\mathcal{O}(L^{-1})\big]
   \\ &=& \ell^{p/2}\int d{\bf q}^{\A}e^{-L [ \epsilon ( {\bf q}^\star )+\mathcal{O}(\alpha)]}\omega(\mathbf{q}^{\B},\sig^\A) = e^{-L\epsilon( {\bf q}^\star)} \times \mathcal{O}(1). \nonumber
\end{eqnarray}
The leading order term of $\langle\ln Z^\A(\mathbf{q}^\star,\sig^\B)\rangle_{\HH}$ is therefore does not involve $\zeta$ and cancels with the leading order term of $L\langle\beta\epsilon(\mathbf{q}(\sig))\rangle_{\HH}$ in \eref{eq:I_genA1} for all $0<\zeta<1$.

The key quantity that depends on $\zeta$ is $Z^{\B}$. Starting from \eref{eq:app_Z2} it can be analyzed similarly to $Z$, by first integrating over the variables $q_2,\dots,q_p$ using the SPA, yielding
\begin{eqnarray}
\fl \quad & Z^{\B}(\mathbf{q}^{\star},\sig^\A) =(L-\ell)^{p/2}\int d{\bf q}^{\B}e^{-(L-\ell)\beta f_{\HH,1-\alpha}(\beta,{\bf q}^{\star},{\bf q}^\B)}\omega(\mathbf{q}^{\B},\sig^\A)
\big[1+\mathcal{O}(L^{-1})\big] \\
 \fl  &=L^{1/2} e^{-L\beta f_\HH (\beta,{\bf q}^\star)} \int dq_{1}^{\B}e^{-L(\det A_{1-\alpha})(q_{1}^{\B})^{2}-L\beta f_{\HH,1-\alpha}^{(4)}(q_{1}^{\B})^{4}+O[(q_{1}^{\B})^{6}]}\tilde{\omega}(q_{1}^{\B},\sig^{\A})\big[1+\mathcal{O}(\ell/L)\big]. \nonumber
\end{eqnarray}
Away from criticality, where $\det A_{1-\alpha}=g(\alpha)=b_0+O(\alpha)$ this integral can be evaluated using a SPA which involve only the first term in the exponential,
yielding $Z^\B \sim O(1)\times e^{-L\beta f_\HH (\beta,{\bf q}^\star)}$, which is similar to the leading order term in $Z$. As a result all the
extensive terms in \eref{eq:I_genA1} cancel, yielding {\it away from criticality} the following expression:
\begin{eqnarray}
 I_{\A:\B}=\mathcal{O}(1) \qquad 0<\zeta\leq 1.
\end{eqnarray}

{\it At criticality}, where $\det A_{1-\alpha}=g(\alpha)=b_1 \alpha+O(\alpha^2)$, one obtains the following expressions for $Z^{\B}$:
\begin{equation}
\label{eq:ZB_finite}
Z^{\B}(\mathbf{q}^{\star},\sig^{\A}) = L^{1/2} e^{-L\beta f_\HH (\beta,{\bf q}^\star)}\int dq_{1}^{\B}e^{-\ell  b_{1} (q_{1}^{\B})^{2}-L(\beta f_{\HH,1-\alpha}^{(4)})(q_{1}^{\B})^{4}+O[(q_{1}^{\B})^{6}]}\tilde{\omega}(q_{1}^{\B},\sig^{\A}) .
\end{equation}
The SPA of the above integral should be evaluated differently depending on the value of $\zeta$. With the appropriate change of variables the leading order term is given by
\begin{equation}
\label{eq:ZB_finite1}
Z^{\B}(\mathbf{q}^{\star},\sig^{\A})\approx e^{-L\beta f_\HH (\beta,{\bf q}^\star)} \tilde{\omega}(q_{1}^\star,\sig^{\A}) \left\{ \begin{array}{ccc}
L^{(1-\zeta)/2}\int dye^{-y^{2}b_{1}} & \quad & \frac{1}{2}<\zeta\leq 1\\
L^{1/4}\int dze^{-z^{2}a \, b_{1}-z^{4}\beta f_{\HH,1-\alpha}^{(4)}} & \quad & \zeta=\frac{1}{2}\\
L^{1/4}\int dze^{-z^{4}\beta f_{\HH,1-\alpha}^{(4)}} & \quad & 0<\zeta<\frac{1}{2}
\end{array}\right. ,
\end{equation}
where  $y\equiv \ell^{1/2}q_1^\B$ and $z\equiv L^{1/4}q_1^\B$. This expression should be compared with that of the partition function $Z=L^{1/4}  e^{-L\beta f_\HH (\beta,{\bf q}^\star)} \times \mathcal{O}(1)$.
For $\zeta > 1/2$, the partition function of subsystem $\B$ is smaller that of the whole system, $Z^\B\ll Z$, whereas for $\zeta \leq 1/2$ they have the same leading order terms.
The effect of subsystem $\A$ on $\B$ becomes negligible in the latter limit. Inserting \eref{eq:ZB_finite1} into \eref{eq:I_genA1} yields in the canonical ensemble the following result {\it at criticality}:
\begin{eqnarray}
I_{\A:\B}=\left\{ \begin{array}{ccc}
(\frac{\zeta}{2}-\frac{1}{4})\ln L+\mathcal{O}(1) & \quad & \frac{1}{2}<\zeta\leq 1\\
\mathcal{O}(1) & \quad & 0<\zeta\leq\frac{1}{2}
\end{array}\right. .
\end{eqnarray}

 The two limits of $\ell\ll L^{1/2}$ and $L^{1/2} \ll \ell \lesssim L$ can be connected smoothly by
 considering $\ell = a L^{1/2}$. The $\mathcal{O}(1)$ part of the mutual information contains in this case a term of the form $\ln [\int dze^{-z^{2}a \, b_{1}-z^{4}\beta f_{\HH,1-\alpha}^{(4)}}]$.
 As expected, this term diverges to $-\infty$ for $a\to\infty$ and converges to the constant term obtained for $\zeta<1/2$ in the $a\to 0$ limit.

The discussion above can be shown to apply also to the micro-canonical ensemble, where obtains the following scaling {\it at criticality}:
\begin{eqnarray}
 I_{\A:\B}=\left\{ \begin{array}{ccc}
( \zeta-\frac{1}{4} ) \ln L+\mathcal{O}(1) & \quad & \ \frac{1}{2}<\zeta\leq 1\\
\frac{\zeta}{2}\ln L+\mathcal{O}(1) & \quad &  0<\zeta\leq\frac{1}{2}
\end{array}\right. .
\end{eqnarray}
As argued in \sref{sec:gen_micro}, the additional $\frac{\zeta}{2}\ln L$ term, which can be written as $\frac{1}{2}\ln \ell$, is due to the fixed energy constraint.

\subsection{Higher order critical points}
\label{sec:higher_order}
Both the BEG model and the NK model exhibit tricritical points, where the second and first order transition lines meet, as shown in \fref{fig:BEG}.
These point are defined by the vanishing of the 4$^{\mathrm{th}}$ order term in the Landau expansion. In the canonical ensemble, we denote the expansion of the Landau free energy
as $ f_\HH(\beta,\tilde{{\bf q}})=f_{\HH,2} q_1^2 + f_{\HH,4} q_1^4+f_{\HH,6} q_1^6+\ldots$
 \footnote{We assumed here that the model is invariant under $q_1\to-q_1$ as in the case when $q_1$ is the overall magnetization in the BEG and NK model.
 This implies that all the odd coefficients in the Landau expansion vanish. Other cases can be treated by following the lines of derivation presented here.}.
 In terms of this expansion the tricritical point is defined by the equation $f_{\HH,2}=f_{\HH,4}=0$.
 As a result, the saddle point approximation performed throughout \sref{sec:generic_model} has to be computed based on the 6$^{\mathrm{th}}$ order terms. For brevity,
 we state only the results of this calculation, which is done by following the same lines of derivation presented above.

We define an $r$-order critical point as such where  the Landau expansion is given by  $f_\HH(\beta,\tilde{{\bf q}})=f_{\HH,2r} q_1^{2r}+\mathcal{O}(q_1^{2r+2})$. For such a general critical point we
obtain the following results in the canonical ensemble:
\begin{eqnarray}
\fl \qquad S_{\A:\B} & =& \gamma_r L^{1-\frac{1}{r}}\left[\alpha^{1-\frac{1}{r}}+(1-\alpha)^{1-\frac{1}{r}}- 1
\right]-\frac{1}{2r}\ln\left[ L \alpha \left( 1-\alpha \right)
\right]+\mathcal{O}\left( 1 \right), \\
\fl \qquad \mathcal{I}_{\A:\B}&=&\frac{1}{2r}\ln L+ \frac{1}{2}\ln\left[ g(\alpha)g(1-\alpha)\right]+
\mathcal{O}\left( 1 \right),
\end{eqnarray}
where $\gamma_r$ denotes a non-universal constant.
At the tricritical point ($r=3$), the separation entropy thus diverges as $L^{2/3}$ as compared to $L^{1/2}$ at the critical point ($r=2$). Similarly, the mutual information
diverges as $\frac{1}{3}\log L$ as compared to $\frac{1}{4}\log L$ at the critical point. This conforms with our intuition that the fluctuation in the order parameter
of the transition are stronger at higher order critical points.

A similar calculation a $r$-order critical point in the micro-canonical ensemble yields
\begin{eqnarray}
\fl \qquad S_{\A:\B} &=& \frac{1}{r}\ln\left[ L \alpha \left( 1-\alpha \right)
\right]+\mathcal{O}\left( 1 \right), \\
\fl \qquad \mathcal{I}_{\A:\B}&=&(\frac{1}{2}+\frac{1}{2r})\ln L+ \frac{1}{2}\ln\left[ g(\alpha)g(1-\alpha)\right]+
\mathcal{O}\left( 1 \right).
\end{eqnarray}
As in the critical case, we obtain an additional logarithmic divergence due to the global energy constraint and no
polynomial divergence in $L$ due to the absence of energy fluctuations.

\section{Conclusions}
\label{sec:conc}
We computed the scaling behavior of two bipartite information estimators, namely the
mutual information and the separation entropy, in a $(p+1)$-state
classical spin chain. The Hamiltonian we considered involves a
mean-field and a short-range interaction term, both of a general form, thus
encompassing a large class of infinite-range interacting models. Two particular examples are the BEG model and NK model.
 Models of this type are particularly interesting because they exhibit a rich
phase diagram while being analytically tractable.

Because mean-field models often exhibit ensemble inequivalence, we chose to study the
information estimators both in the canonical and in the micro-canonical ensemble. We first studied the
limit where the system is fictitiously divided into two macroscopic subsystems, with total number of spins $\ell$ and $L-\ell$, respectively.
In the canonical ensemble and away from the critical line , both the estimators remain finite in the
thermodynamic limit. However, at criticality, this ceases to be true. The
mutual information diverges as $(1/4)\ln L$ where the pre-factor $1/4$ appears to
be a characteristic of the mean-field interaction. The coefficient is universal
in the sense that it does not depend on the microscopic details of the model, and thus remains constant along the critical line.
This is not true for the scaling function of the mutual information, which depends on the details of the model.
The separation entropy at criticality exhibits a
$\sqrt{L}$ divergence, whose coefficient in general is non-universal. However, the sub-leading term diverges logarithmically with $L$ and exhibits a universal scaling form $(1/4)\ln
\left[L \alpha (1-\alpha)\right]$, where $\alpha \equiv \ell/L$.
It is important to note that unlike in the entanglement entropy used for quantum systems, the coefficient of the $\ln L$ term in the two estimators does not depend on the number of states each spin takes.

In the micro-canonical ensemble we find a different scaling behaviour.
Both of the estimators exhibit $\ln L$ corrections to the area law even away from the critical line. This
divergence is argued to be due to the fixed total energy constraint in
the microcanonical ensemble, which leads to non-trivial correlation among all spins.
This difference between the canonical and the micro-canonical calculations is not related to the ensemble inequivalence observed in
long-range interacting systems. The same difference is observed in spin-chains without interactions or with only short-range interactions. This can be easily seen by dropping the mean-field interactions term from our analysis.

The universal scaling of the shared-information estimators opens a new direction
in analyzing critical phenomena in classical systems. They could be useful in detecting phase transitions and in identifying
universality classes. In this paper we find universal scaling behavior of the bipartite information
estimators in a one-dimensional mean-field spin
chain. It would be interesting to investigate how the scaling behavior changes in
higher-dimensions, and in continuous models. A few promising non-trivial models to
analyze are the classical two-dimensional Ising model and ice-type models, where
many beautiful exact results are available in the literature.

Another interesting direction would be to study the information estimator in classical
non-equilibrium spin-chains, where equilibrium concepts such as free energy cannot be used to analyze critical phenomena.
 Non-equilibrium systems are particularly relevant
for our study, because similar to the mean-field model they generally exhibit
long-range correlations \cite{Spohn1983,Sadhu2014}, ensemble inequivalence
\cite{Cohen2012} and phase transitions in one-dimension \cite{David2000}. It
would be interesting to know how the scaling behavior of the information estimators in these systems compare with the results of our present study.

\ack We thank F.C. Alcaraz, H. Hinrichsen, T. Mori and D. Mukamel for helpful discussions. The support of the Israel Science
Foundation (ISF) is gratefully acknowledged.

\appendix

\section{Vanishing of the $\sqrt{L}$ term in the mutual information}
\label{app:vanish}
In this appendix it is shown that the mutual information estimator does
not exhibit a $\sqrt{L}$-divergence at criticality for the generic mean-field model studied \sref{sec:generic_model}.
In order to understand the source of $\sqrt{L}$-divergence in the mutual information, it is useful to consider the thermodynamic average of a general extensive function,
$L g({\bf Q}(\sig)/L)$. Following the same derivation that leads to the expression of $S$ in \eref{eq:S_int} yields a similar expression for the average over $g$,
\begin{equation}
\label{eq:appB_generalG}
\langle L g({\bf q}(\sig)) \rangle_\HH  =  Z^{-1}L^{p/2}\int d{\bf q} \omega(\mathbf{q}) L g({\bf q}) e^{-L\beta f_\HH(\beta,{\bf q})}\big[1+\mathcal{O}(L^{-1})\big].
\end{equation}
As in the calculation of $S$, the above integral can be evaluated by a two step SPA. In the first step the integrals over $q_2,\dots,q_p$ are evaluated, yielding
\begin{eqnarray}
\label{eq:appB_generalG1}
\langle L g({\bf q}(\sig)) \rangle_\HH  & = &  \frac{\int dm \left[ L g(\tilde{\bf q}(q_1)) \right] \tilde{\omega}(q_1) e^{-L f_\HH(\tilde{\bf q}(q_1))}}{\int dm  \tilde{\omega}(\tilde{\bf q}(q_1)) e^{-Lf_\HH(\tilde{\bf q}(q_1))}} +\mathcal{O}\left( 1 \right).
\end{eqnarray}
In order to perform the SPA of the remaining integral over $q_1$ one has to expand $g(\tilde{\bf q}(q_1))\tilde{\omega}(q_1)$, $\tilde{\omega}(q_1)$ and $f_\HH(\tilde{\bf q}(q_1))$ in powers of
$q_1$. For brevity, it is useful to write the expansions using the following notation:
\begin{equation}
\psi^{(s)}_\circ=\frac{1}{s!}\frac{d^{s} \psi(\tilde{\bf q}(q_1))}{d q_1^s}\Big\vert_{q_1^\star},
\end{equation}
where $\psi$ denotes a general function.

At criticality, the SPA of the integrals over $q_1$ in \eref{eq:appB_generalG1} yields
\begin{eqnarray}
\label{eq:app_g1}
\langle L g({\bf q}(\sig))\rangle_\HH
 & = & \frac{ \int dq_1  L \left[ \tilde{\omega}_\circ g_\circ^{(0)}  +  (\tilde{\omega} g)^{(2)}_\circ (q_1-q_1^\star)^2 \right] e^{-L f^{(4)}_\circ (q_1-q_1^\star)^4}}
{\int dq_1 \left[ \tilde{\omega}_\circ +  \tilde{\omega}^{(2)}_\circ (q_1-q_1^\star)^2 \right] e^{-L f^{(4)}_\circ (q_1-q_1^\star)^4}} +\mathcal{O}\left( 1 \right) \nonumber \\
& = & L g_\circ^{(0)} + \sqrt{L} \left(\tilde{\omega}^{(1)}_\circ g^{(1)}_\circ+g^{(2)}_\circ\right) \frac{\chi_{2,4}\left(f^{(4)}_\circ \right)}{\chi_{0,4}\left(f^{(4)}_\circ \right)}
+\mathcal{O}\left( 1 \right) .
\end{eqnarray}
Note that above we assumed that $f^{(2)}_\circ=0$ since the system is critical.

\Eref{eq:app_g1} implies that the $\mathcal{O}(\sqrt{L})$ term depends on the derivatives of $g$ with respect to the order parameter.
Specifically, when $g$ is chosen to be the Landau free energy, the $\mathcal{O}(\sqrt{L})$ term vanishes because $f^{(1)}_\circ=f^{(2)}_\circ=0$ at criticality.
As a result the correction term to the average of $L f({\bf q}(\sig))$ is independent of $L$ and is given by
\begin{equation}
\label{eq:app_f1}
\langle L f_\HH({\bf q}(\sig)) \rangle_\HH = \frac{L^{p/2}}{Z}\int d{\bf q} \omega(\mathbf{q}) L f_\HH({\bf q}) e^{-L\beta f_\HH(\beta,{\bf q})}\big[1+\mathcal{O}(L^{-1})\big] = L f_\HH({\bf q}^\star) +\mathcal{O}\left( 1 \right).
\end{equation}
In fact, the above scaling is true also away from criticality, where $f^{(2)}_\circ \neq 0$. In this case the second order term in the SPA
of the integrals over $q_1$ in \eref{eq:appB_generalG1} scales as $\mathcal{O}(1)$. This leads in term to an $\mathcal{O}\left( 1 \right)$ correction term
to $\langle L f({\bf q}(\sig))\rangle_\HH$ as found above at criticality.
In the following we demonstrate using \eref{eq:app_f1} that the mutual information does exhibit an $O(\sqrt{L})$ divergence.

As shown in \eref{eq:I_genA} the mutual information estimator can be written for the generic mean-field model in the following form:
\begin{equation}
I_{\A:\B}=\ln Z+\big\langle\ln[Z^{\A}(\mathbf{q}({\sig}^{\B}),\sig^\B)Z^{\B}(\mathbf{q}({\sig}^{\A}),{\sig}^{\A})]\big\rangle_{\HH}-L\big\langle\beta \epsilon(\mathbf{q}(\sig))\big\rangle_{\HH}+\mathcal{O}(1).
\end{equation}
By adding and subtracting the an average over the free energy of the short-range interacting model of each subsystems,
 $\langle (1-\alpha) \beta f_\phi({\bf q}(\sig^\B))  \rangle_\HH$ and $\langle \alpha \beta f_\phi({\bf q}(\sig^\A))  \rangle_\HH$,
 the above equation can be written as
\begin{eqnarray}
\label{eq:app_I1}
\fl \qquad &I_{\A:\B}=&\ln Z+\big\langle \ln
Z^{\A}(\mathbf{q}({\sig}^{\B}),\sig^\B)+(1-\alpha) \beta f_\phi({\bf
q}(\sig^\B))  \big\rangle_\HH \nonumber\\
\fl \qquad & & + \big\langle \ln Z^{\B}(\mathbf{q}({\sig}^{\A}),\sig^\A)+\alpha
\beta f_\phi({\bf q}(\sig^\A))  \big\rangle_\HH\nonumber\\
&& -L\big\langle\beta f_\HH(\beta,\mathbf{q}(\sig^\A),\mathbf{q}(\sig^\B))\big\rangle_{\HH}+\mathcal{O}(1).
\end{eqnarray}
In the following, we demonstrate that each of the above three expectation values is of the form of \eref{eq:app_f1} and therefore does not yield a $\sqrt{L}$ divergence.
This is done using the expression for the average of a general function of ${\bf q}({\sig}^{\A})$ and ${\bf q}({\sig}^{\B})$,
\begin{eqnarray}
\label{eq:app_gqaqb}
\big\langle g({\bf q}({\sig}^{\A}),{\bf q}({\sig}^{\B})) \big\rangle_\HH  =
Z^{-1} \ell^{\frac{p}{2}}(L-\ell)^{\frac{p}{2}}\nonumber\\
\int d{\bf q}^{\A} d{\bf q}^{\B}g(\mathbf{q}^{\A},\mathbf{q}^{\B}) \omega(\mathbf{q}^{\A},\mathbf{q}^{\B}) e^{-L\beta f_{\HH}(\beta,\mathbf{q}^{\A},\mathbf{q}^{\B})}  [1+\mathcal{O}(L^{-1})],
\end{eqnarray}
obtained in \eref{eq:avg_qa_qb} and repeated here for completeness. This implies immediately that, the third expectation value in \eref{eq:app_I1} is of the form of  \eref{eq:app_f1}
and hence $\big\langle\beta f_\HH(\beta,\mathbf{q}(\sig^\A),\mathbf{q}(\sig^\B))\big\rangle_{\HH} = f_\HH(\beta,\mathbf{q}^\star,\mathbf{q}^\star)+\mathcal{O}(1)$.

The two other expectation values in \eref{eq:app_I1} can be treated in a similar manner. For $Z^\B$ this is done by evaluating the following integral:
\begin{equation}
\label{eq:app_ZBc}
 Z^{\B}=(L-\ell)^{p/2}\int d{\bf q}^{\B}e^{-(L-\ell)\beta f_{\HH,1-\alpha}(\beta,{\bf q}^\A,{\bf q}^\B)}\omega(\mathbf{q}^{\B},\sig^\A)
\big[1+\mathcal{O}(L^{-1})\big],
\end{equation}
obtained in \eref{eq:app_PMA}.
The SPA of the integrals over ${\bf q}^\B$ in \eref{eq:app_ZBc} yields
\begin{eqnarray}
\label{eq:app_ZBd}
 \ln Z^{\B}(\mathbf{q}^{\A},\sig^\A)= -(L-\ell)\beta f_{\HH,1-\alpha}(\beta,{\bf q}^\A,{\bf q}^{\star \,\B}({\bf q}^\A))+ \mathcal{O}(1),
\end{eqnarray}
where ${\bf q}^{\star \,\B}({\bf q}^\A)$ denotes the saddle point as a function of ${\bf q}^\A$. It is important to note that since the boundary interaction term with
subsystem $\A$ appears only in $\omega(\mathbf{q}^{\B},\sig^\A)$ in \eref{eq:app_ZBc} it does not affect the value of the saddle point.
When considering a function only of ${\bf q}({\sig}^{\A})$ in \eref{eq:app_gqaqb}, one obtains
\begin{equation}
\label{eq:app_gqa}
\big\langle g({\bf q}({\sig}^{\A})) \big\rangle_\HH  = Z^{-1} \ell^{\frac{p}{2}}\int d{\bf q}^{\A} g(\mathbf{q}^{\A}) e^{
-L\alpha \beta f_\phi({\bf q}^\A) -(L-\ell)\beta f_{\HH,1-\alpha}(\beta,{\bf q}^\A,{\bf q}^{\star \,\B}({\bf q}^\A)) + \mathcal{O}(1) } [1+\mathcal{O}(L^{-1})].
\end{equation}
Combing \eref{eq:app_ZBd} and \eref{eq:app_gqa} implies that, the second expectation value in \eref{eq:app_I1} is also of the form of  \eref{eq:app_f1} and
hence
\begin{equation}
\big\langle \ln Z^{\B}(\mathbf{q}({\sig}^{\A}),\sig^\A)+\alpha \beta f_\phi({\bf q}(\sig^\A))  \big\rangle_\HH = L [ \alpha \beta f_\phi({\bf q}^\star) + (1-\alpha)
f_{\HH,1-\alpha}(\beta,{\bf q}^\star,{\bf q}^\star)] +\mathcal{O}(1)
\end{equation}
Clearly the same derivation can be used to show that the first expectation value in \eref{eq:app_I1} does not exhibit a $\sqrt{L}$ divergence.

In addition to the absence of the $\mathcal{O}(\sqrt{L})$, the leading $\mathcal{O}(L)$ terms in the three expectation values can be shown to vanish yielding
\begin{eqnarray}
\label{eq:app_I2}
I_{\A:\B} = \ln Z + \mathcal{O}(1),
\end{eqnarray}
both at criticality and away from criticality. It is important to note that in the above analysis we did
not consider the dependence on $\ln \alpha$ explicitly. This dependence is studied in \sref{sec:gen_can} in the canonical ensemble.

\pagebreak

\bibliographystyle{unsrt}
\bibliography{reference}

\end{document}